\documentclass[useAMS,usenatbib,onecolumn]{mn2e}

\usepackage{amsfonts,amsmath,amssymb,mathrsfs}
\usepackage{graphicx,epsf,epsfig}
\usepackage{bm}
\usepackage{longtable}
\usepackage[usenames,dvipsnames]{xcolor}
\usepackage{multirow}

\usepackage{graphicx}
\usepackage{times}
\usepackage{color}
\usepackage{breakurl}
\usepackage{color}
\usepackage{mathrsfs}

\usepackage{hyperref}
\hypersetup{
  colorlinks=true,        
  linkcolor=blue,         
  citecolor=cyan,         
}

\newcommand{\cf}{cf.,~}
\newcommand{\ie}{i.e.,~}
\newcommand{\eg}{e.g.,~}

\title[EM fields in exterior of relativistic star -- II]
      {Electromagnetic fields in the exterior of an oscillating
        relativistic star -- II. Electromagnetic damping}

\author[L.~Rezzolla and B.~J.~Ahmedov]{Luciano~Rezzolla$^{1,\;2}$ and
  Bobomurat~J.~Ahmedov$^{3,\;4,\;5}$ \\
$^{1}$Institute for Theoretical Physics, Max-von-Laue-Str. 1, D-60438
  Frankfurt, Germany \\
$^{2}$Frankfurt Institute for Advanced Studies, Ruth-Moufang-Str. 1,
  D-60438 Frankfurt, Germany \\
$^{3}$Institute of Nuclear Physics, Ulughbek, Tashkent 100214,
  Uzbekistan\\
$^{4}$Ulugh Beg Astronomical Institute, Astronomicheskaya 33, Tashkent
  100052, Uzbekistan \\
$^{5}$National University of Uzbekistan, Tashkent 100174, Uzbekistan \\
}
\pagerange{\pageref{firstpage}--\pageref{lastpage}}

\pubyear{2016}

\begin{document}

\maketitle

\label{firstpage}

\begin{abstract}
    An important issue in the asteroseismology of compact and magnetized
    stars is the determination of the dissipation mechanism which is most
    efficient in damping the oscillations when these are produced. In a
    linear regime and for low-multipolarity modes these mechanisms are
    confined to either gravitational-wave or electromagnetic losses. We
    here consider the latter and compute the energy losses in the form of
    Poynting fluxes, Joule heating and Ohmic dissipation in a
    relativistic oscillating spherical star with a dipolar magnetic field
    in vacuum. While this approach is not particularly realistic for
    rapidly rotating stars, it has the advantage that it is fully
    analytic and that it provides expressions for the electric and
    magnetic fields produced by the most common modes of oscillation both
    in the vicinity of the star and far away from it. In this way we
    revisit and extend to a relativistic context the classical estimates
    of McDermott et al. Overall, we find that general-relativistic
    corrections lead to electromagnetic damping time-scales that are at
    least one order of magnitude smaller than in Newtonian
    gravity. Furthermore, with the only exception of $g$ (gravity) modes,
    we find that $f$ (fundamental), $p$ (pressure), $i$ (interface) and
    $s$ (shear) modes are suppressed more efficiently by gravitational
    losses than by electromagnetic ones.
\end{abstract}



\section{Introduction}
\label{introd}

Neutron stars are endowed with intense electromagnetic fields, but they
may also subject to oscillations of various type, such as those observed
as quasi-periodic oscillations (QPOs) following giant flares of soft
gamma-ray repeaters \citep[SGRs;][]{Israel2005, Strohmayer2005,
Strohmayer2006, Watts2006, Watts07a, Watts07b, Huppenkothen2014a,
  Turolla2015}. The analysis of the X-ray data from SGRs has in fact
revealed that the decaying part of spectrum exhibits a number of
oscillations with frequencies in the range of a few tenths of Hz to a few
hundreds Hz, that agree reasonably well with the expected toroidal modes
of the magnetar crust \citep{Duncan1998}. For example, QPO frequencies
around 18, 26, 30, 92, 150, 625, and 1840 Hz were discovered in the
outburst of SGR1806-20\footnote{Using a different technique,
  \citet{Hambaryan2011} have also found new QPOs in the data of the
  SGR1806-20 giant burst at frequencies 16.9, 21.4, 36.4, 59.0 and 116.3
  Hz.}, while QPO frequencies around 28, 53, 84 and 155 Hz were detected
for the outburst of SGR1900+14. More recently, \citet{El-Mezeini2010}
claimed the discovery of 84, 103 and 648 Hz QPOs in the normal bursts of
SGR1806-20. Besides providing the first evidence for oscillations in
compact stars, these observations are very important because they
motivate further study in neutron star oscillation modes as a powerful
tool for probing the complex stellar interior.

The attempts to explain the above QPO frequencies in SGRs by using the
crust toroidal oscillations has produced a vast literature in a very
short time \citep{Levin2006, Glampedakis2006, Samuelsson2007, Sotani2007,
  Levin2007, Sotani2008, Colaiuda2009, Cerda2009, Sotani2009, Gabler2011,
  Gabler2012, Gabler2013b, Gabler2014}. Overall, all of these works have
shown that an Alfv\'en QPO model is a viable explanation for the observed
QPOs if a number of suitable conditions are met by the magnetic-field
strength and geometry. Additional observations are clearly needed to
impose additional constraints on the different models. At the same time,
a considerable effort has also been dedicated to the establishing how
much of the vibrational modes excited in magnetar giant flares can lead
to the excitation of $f$ modes and to the copious emission of
gravitational waves. Also in this case, the various works have pointed
out that only a small fraction of the flare's energy is converted
directly into the lowest order $f$ modes and that the corresponding
gravitational-wave emission is small
\citep{Ciolfi2011,Lasky2011,Ciolfi2012,Zink2012}.

Crustal oscillations of nonrotating non-magnetized neutron stars have
been studied in Newtonian theory \citep{McDermott1985, McDermott1988} as
well as in general relativity \citep[\eg][]{Schumaker1983, Finn1990}. The
study of the oscillations of magnetized and rotating relativistic stars
is far from being trivial or fully solved. At least in principle, in
fact, it involves the solution of the coupled Einstein-Euler-Maxwell
equations, as well as a detailed modelling of the stellar structure via
realistic equations of state. In practice, however, the investigations
have so far concentrated on two distinct but complementary
approaches. The first one has focused on the study of the
eigenfrequencies and eigenfunctions of selected modes of magnetized
neutron stars in general relativity \citep[\eg][]{{Messios2001},
  {Chugunov2005}, Piro2005, Sotani2007, {Glampedakis2006},
  {Samuelsson2007},{Vavoulidis2007}}]. The second one, instead, has
  focused its attention on the properties of the electromagnetic fields
  from near the stellar surface up to the wave-zone
  \citep[\eg][]{McDermott1984, {Muslimov1986},
    {Muslimov1997}, {Konno1999}, {Konno2000a}, {Konno2000b}, {Timokhin2000}, {Rezzolla2001}, {Rezzolla2001b},
    {Kojima2004b}, {Rezzolla2004}}\footnote{It
    is important to clarify that the solutions presented in
    \citet{Rezzolla2001b, Rezzolla2001} are exact in the slow-rotation
    approximation; see also \citet{Petri2013} for a recent re-derivation
    of the very same equations using a different approach.}. The work
  reported in this paper belongs to the second class of studies and
  concentrates on calculating explicit expressions for the
  electromagnetic fields generated by radial, toroidal and spheroidal
  oscillations of the stellar surface.

A number of different motivations are behind this investigation.  First,
we intend to provide analytical expressions for the electric and magnetic
fields produced by the most common modes of oscillation for a
relativistic magnetized spherical star in vacuum, both in the vicinity of
the star and far away from it. In this way we specialize the generic
expressions for the vacuum electromagnetic fields produced by a
relativistic magnetized spherical star presented in \citet{Rezzolla2004}
to the most common oscillation modes. Secondly, we attempt to address an
important issue in the asteroseismology of compact and magnetized
relativistic stars, namely, the general-relativistic corrections on the
dissipation mechanisms that are most efficient in damping the
oscillations. Because in a linear regime, these mechanisms are confined to
either gravitational-wave or electromagnetic losses, here we compute and
compare, within a general-relativistic framework, the energy losses as
produced by the oscillations in the form of Poynting fluxes, Joule
heating and by the gravitational-wave emission. In this way, we revisit
and extend to a general-relativistic context the Newtonian estimates of
\citet{McDermott1988} for the damping times of oscillating magnetized
neutron stars due to electromagnetic radiation.

Overall, a number of factors, such as the type of mode, the
magnetic-field strength and the compactness of the star, concur in
determining what is the main damping mechanism of the
oscillations. However, we have found that the following results are
generically true for a typical neutron star with a dipolar magnetic field
of $\sim 10^{12} \,{\rm G}$: \textit{(i)} the general-relativistic
corrections to the electromagnetic fields lead to damping time-scales due
to electromagnetic losses which are at least one order of magnitude
smaller than their Newtonian counterparts; \textit{(ii)} with the only
exception of $g$ (gravity) modes, we find that $f$ (fundamental), $p$
(pressure), $i$ (interface) and $s$ (shear) modes are suppressed more
efficiently by gravitational losses than by electromagnetic ones;
\textit{(iii)} Joule heating is not as an important damping mechanism in
general relativity as it is in Newtonian gravity.

The paper is organized according to the following plan: in
Section~\ref{em_nz} we provide a detailed discussion of the
electromagnetic fields produced by spheroidal and toroidal oscillations
in the vicinity of the stellar surface (\ie in the 'near zone'), while
the following Section~\ref{em_wz} is devoted to the calculation of the
corresponding fields at a large distance from the star (\ie in the 'wave
zone'). In both cases, we show that in the Newtonian limit our solutions
coincide with those obtained by \citet{Muslimov1986}. In
Section~\ref{oscilltn_damping} we instead investigate the damping due to
electromagnetic emission when the star is modelled as a relativistic
polytrope with infinite conductivity, as well as the damping due to Joule
heating when the stellar matter has high but finite conductivity. The
damping times are then compared with the corresponding damping times due
to gravitational-wave emission. In spite of its general-relativistic
amplification, Ohmic dissipation is shown to be almost insignificant
because of the small relative motions of the stellar crust matter with
respect to the field lines due to the high electric conductivity in
neutron stars. Finally, Section~\ref{conclusions} collects our results
and discusses the prospects of future work.

We use in this paper a system of units in which $c = 1$, a space-like
signature $(-,+,+,+)$ and a spherical coordinate system $(t,r,\theta
,\phi)$. Greek indices are taken to run from 0 to 3, Latin indices from 1
to 3 and we adopt the standard convention for the summation over repeated
indices. We will indicate four-vectors with bold symbols (\eg
$\boldsymbol{u}$) and three-vectors with an arrow (\eg
$\vec{\boldsymbol{u}}$).

\section{Electromagnetic fields in the near-zone}
\label{em_nz}

The study of the electromagnetic fields produced by an oscillating and
magnetized neutron star has a long history and a first comprehensive
investigation was carried out by \citet{McDermott1984}, who estimated the
emission of electromagnetic radiation in the wave zone from nonradial
pulsations of a neutron star with a strong dipolar magnetic field in
vacuum, and who provided the first estimates of the electromagnetic
damping within a Newtonian description of gravity.  Subsequently,
\citet{Muslimov1986} derived, again in Newtonian gravity, exact
analytical solutions for the vacuum electromagnetic fields around an
oscillating neutron star with a dipolar surface magnetic field (both in
the near and wave zones), while \citet{McDermott1988} completed a
detailed analysis of various oscillation modes and computed the
corresponding damping times. Later on, \citet{Duncan1998} improved the
results of \citet{McDermott1988} with better estimates for the parameters
of typical neutron stars, and has calculated the eigenfrequencies of the
toroidal modes of a magnetar.

At about the same time, \citet{Timokhin2000} have pointed out another
interesting aspect of oscillating magnetized stars and, in particular,
that a neutron star with magnetic field $B$ and oscillating at a
frequency $\omega$ will generate an electric field $E\sim \omega\xi B/c
$, where $\xi$ is the characteristic (linear) amplitude of the
oscillations. Interestingly, such an electric field can be strong enough
to pull charged particles from the surface and create a magnetosphere
even in the absence of rotation. These results were first extended to the
general-relativistic context for a relativistic spherical nonrotating
neutron star by \citet{Abdikamalov2009}. In \citet{Morozova2010},
instead, it has been shown that the electromagnetic energy losses from
the polar cap region of a rotating neutron star can be significantly
enhanced if oscillations are also present, and, for the mode
spherical-harmonics indices $(\ell,m)=(2,1)$, such electromagnetic losses
are a factor of $\sim 8$ larger than the rotational energy losses, even for
a velocity oscillation amplitude at the star surface as small as
$\tilde\eta=0.05 \ \Omega \ R$, where $\Omega$ { and $R$ are the angular
  velocity and the radius of the neutron star, respectively.}  In
\citet{Morozova2012}, on the other hand, the conditions for radio
emission in magnetars have been considered and it has been found that,
when oscillations of the magnetar are taken into account, the radio
emission from the magnetosphere is generally favoured. Indeed, the major
effect of the oscillations is to amplify the electric potential in the
polar cap region of the magnetar magnetosphere.

In \citet{Rezzolla2004} (hereafter paper I), we have derived exact
analytical expressions for the interior and exterior electromagnetic
fields in a perfectly conducting relativistic star, expressing them in
terms of generic velocity and magnetic fields. Starting from that work,
we here specialize those expressions to the most common modes of
oscillations. For simplicity we will assume star to be in vacuum and
refer the discussion of the case in which a magnetosphere is present to
the works of \citep{Abdikamalov2009, {Morozova2010}, {Morozova2012},
  Zanotti2012, {Morozova2014}, {Lin2015}}.

Because our analysis is essentially analytical, we simplify the treatment
by considering separately the electromagnetic fields in the vicinity of
the stellar surface, \ie in the 'near zone', where they are almost
stationary, and in regions far away from the surface, \ie in the 'wave
zone', where they assume the properties of electromagnetic radiation. In
doing this we will assume the space-time to be that of a spherical
relativistic star of mass $M$ with a line element that in a spherical
coordinate system $(t,r,\theta ,\phi)$ is given by
\begin{equation}
\label{schw}
d s^2 = g_{00} {\rm d} t^2 + g_{11} {\rm d} r^2+
        r^2 ({\rm d} \theta ^2+ \sin^2\theta {\rm d} \phi ^2) \,.
\end{equation}
The portion of the space-time exterior to the star (\ie for $r \ge R$) is
simply given by the Schwarzschild solution with $-g_{00} = N^2 := (1 -
2M/r)$, $g_{11} = -1/g_{00}$, and where $N_{_\textrm{R}}^2 := 1 - {2M}/{R}$ is the
redshift at the stellar surface.

In what follows, we will first discuss the generic form of the magnetic
and electric fields in the near zone and then illustrate the expressions
they assume in the case of radial, spheroidal and toroidal oscillations,
respectively. We will then present the generic expressions for the
perturbed magnetic and electric fields, and, subsequently proceed to
providing the corresponding expressions for most common modes of
oscillations.

\subsection{General expressions for the magnetic field}

As shown in paper I, given a background magnetic field decomposed in
terms of spherical harmonics [see equations 55--57 of paper I], it is
possible to express the time dependence of a linear perturbation in the
magnetic field components ${\delta B}^{\hat i}$ in terms of new
functions, $\delta s_{\ell m}(t)$, which are linear superposition of the
background ones. As a result, according to equations 67--69 of paper I,
when a generic velocity perturbation of the type
\begin{equation}
\label{vel} \delta u^\alpha\vert_{r=R} = \frac{1}{N_{_\textrm{R}}}
\bigg(1,
    \delta v^i_{_\textrm{R}}\bigg)=
       \frac{1}{N_{_\textrm{R}}} \bigg(1,N_{_\textrm{R}}\,\delta v_{_\textrm{R}}^{\hat r},
    \frac{\delta v_{_\textrm{R}}^{\hat\theta}}{R},
        \frac{\delta v_{_\textrm{R}}^{\hat\phi}}{R\sin\theta}\bigg) \,,
\end{equation}
and
\begin{equation}
\delta u_\alpha\vert_{r=R} =
        \frac{1}{N_{_\textrm{R}}} \bigg(-N_{_\textrm{R}}^2,
        \frac{\delta v_{_\textrm{R}}^{\hat r}}{N_{_\textrm{R}}},
    {R \,\delta v_{_\textrm{R}}^{\hat\theta}},
        {R\sin\theta \,\delta v_{_\textrm{R}}^{\hat\phi}}\bigg) \,,
\end{equation}
(where $\delta v_{_\textrm{R}}^i:= {\rm d}x^i/{\rm
  d}t\vert_{r=\textrm{R}}$ is the oscillation three-velocity of the
conducting stellar medium at the stellar surface)
is introduced over a background dipolar magnetic field (\ie with
$\ell=1$), the complete expressions for the newly generated
magnetic-field components are given by the real parts of the following
complex expressions
\begin{eqnarray}
\label{nz_mfg1}
&& \delta B^{\hat r} = \frac{1}{M^3}\left[\ln N^2+
    \frac{2M}{r}\left(1+\frac{M}{r}\right)\right]
    \delta s_{1m} \,{\rm e}^{{\rm i}m\phi} \cos\theta \,, \\
\nonumber \\
\label{nz_mfg2} && \delta B^{\hat \theta} =
- \frac{N}{M^2r}\left(
        \frac{r}{M}\ln N^2+
    \frac{1}{N^2}+1\right)
     \delta s_{1m}
    \,{\rm e}^{{\rm i}m\phi} \sin\theta\,, \\
\nonumber \\
\label{nz_mfg3} && \delta B^{\hat \phi} = {\rm i}m
\frac{N}{M^2r\sin\theta}
    \left(\frac{r}{M}\ln N^2+
    \frac{1}{N^2}+1\right) \delta s_{1m}
    \,{\rm e}^{{\rm i}m\phi} \cos\theta \,,
\end{eqnarray}
where $A^{\hat i}:= \boldsymbol{\tilde \omega}^{\hat i}_{\textrm{k}} A^k$
are the components of an arbitrary quantity $A^k$ in the orthonormal
frame carried by static observers [see equations 6--9 of paper I for the
  explicit expressions of the 1-forms $\boldsymbol{\tilde\omega}^{\hat
    i}_{\textrm{k}}$]. The values of the integration constants $\delta
s_{1m}(t)$ can be calculated rather straightforwardly if the oscillation
modes are assumed to have a harmonic time dependence of the type ${\rm
  exp}(-{\rm i}\omega_{_\textrm{R}} t)$, where $\omega_{_\textrm{R}}$ is
the mode frequency at the stellar surface and corresponds to the real
part of the complex mode eigenfrequency. In this case, using equation
(B4) from paper I and thus after requiring the continuity of the
tangential electric field at the stellar surface, we obtain the following
condition for the integration constant
\begin{eqnarray}
\label{coef_perturb}
&& \partial_{\rm t} \delta s_{1m}(t)\vert_{r=R}=
    -\frac{3R}{8f_{_\textrm{R}}}
    \int {\rm d}\Omega Y^*_{1 m}\Bigg\{
    (\nabla^2_{_{\Omega}} S) \frac{1}{R\sin\theta}\left[
    \partial_{\rm \theta} \left(\sin\theta \,\delta v^{\hat\theta}\right) +
    \partial_{\rm \phi} \,\delta v^{\hat \phi}\right]
    +N_{_\textrm{R}}\left[(\partial^2_{\theta r} S)\,\partial_{\rm \theta}
    \delta v^{\hat r}+
    \frac{1}{\sin^2\theta}(\partial^2_{\phi r} S)
    \,\partial_{\rm \phi} \delta v^{\hat r}\right]+
\nonumber\\\nonumber\\
&&
\hskip 2.5 cm
    \left[N_{_\textrm{R}}\,\partial_{\rm r}
    \left(\nabla^2_{_{\Omega}} S\right)\,\delta v^{\hat r}
    +\frac{1}{R}\,\partial_{\rm \theta}\left(\nabla^2_{_{\Omega}} S\right)
    \,\delta v^{\hat\theta}
    + \frac{1}{R\sin\theta}\,\partial_{\rm \phi}
    \left(\nabla^2_{_{\Omega}} S\right)\,\delta v^{\hat\phi}\right]
    \Bigg\}\Bigg\vert_{r=R}\,,
\end{eqnarray}
where $S=S_{1m}(r)Y_{1m}$, ${\rm d}\Omega=\sin\theta {\rm d}\theta {\rm d}\phi$, and
$\nabla^2_{_{\Omega}}$ is the angular part of the Laplacian, \ie
\begin{equation}
\nabla^2_{_{\Omega}}:=\frac{1}{\sin\theta}\,\partial_{\rm \theta}
\left(\sin\theta\,\partial_{\rm \theta}\right)+
\frac{1}{\sin^2\theta}\partial^2_{\phi}\,.
\end{equation}
Note that differently from what done in paper I, here the integration
constants have acquired also a time dependence as a result of the
dynamics of the stellar surface across which the boundary conditions need
to be imposed. However, as it will be shown in Section~\ref{spheroid},
these corrections are negligible at first order in the perturbation.

For a dipolar magnetic field and using equation 53 of paper I, the
'magnetic' scalar function $S$ in equation (\ref{coef_perturb}) is given
by
\begin{equation}
\label{S1m}
S = \frac{r^2}{2M^3}\left[\ln N^2+
    \frac{2M}{r}\left(1+\frac{M}{r}\right)\right]
    s_{1m}\,Y_{1m}\,,
\end{equation}
where $Y_{\ell m}$ are the standard spherical harmonics and the first
$s_{1m}$ coefficients are given by
\begin{equation}
\label{slm} s_{10}=-\frac{\sqrt{3\rm{\pi}}}{4} B R^3\cos\chi \,, \hskip
2.0 cm s_{11}= \sqrt{\frac{3\rm{\pi}}{2}} \frac{B R^3}{2}\sin\chi\,.
\end{equation}
Here $B := 2 \mu/R^3$ is the (Newtonian) value of the surface dipolar
magnetic field at the magnetic pole and $\mu$ is magnetic moment, $\chi$
is the inclination angle between the magnetic axis and the spherical
polar axis. As a result, the integration constants defined in equation
(\ref{coef_perturb}) have the explicit expressions
\begin{eqnarray}
\label{coef_perturb_modif}
&& \partial_{\rm t} \delta s_{1m}(t)\vert_{r=R}=
    \frac{3B R^2}{8f_{_\textrm{R}}}
    \int {\rm d}\Omega Y^*_{1 m}\Bigg\{
    (\cos\chi \cos\theta +
        \sin\chi \sin\theta \,{\rm e}^{{\rm i}\phi})
    \left[\frac{f_\textrm{R}}{\sin\theta}\left[
    \partial_{\rm \theta} \left(\sin\theta \,\delta v^{\hat\theta}\right) +
    \partial_{\rm \phi} \,\delta v^{\hat \phi}\right]- 2h_{_\textrm{R}}\,\delta v^{\hat
    r}\right]-
\nonumber\\\nonumber\\
&&
\hskip 2.5 cm
    (\cos\chi \sin\theta
    - \sin\chi \cos\theta \,{\rm e}^{{\rm i}\phi})\left[h_{_\textrm{R}}\,\partial_{\rm \theta}
    \,\delta v^{\hat r}+ f_{_\textrm{R}}\,\delta v^{\hat\theta}\right]+
    {\rm i}\sin\chi \,{\rm e}^{{\rm i}\phi}
     \left[f_{_\textrm{R}}\,\delta v^{\hat\phi}+
    \frac{1}{\sin\theta} h_{_\textrm{R}}
    \,\partial_{\rm \phi} \delta v^{\hat r}\right]
    \Bigg\} \,.
\end{eqnarray}
In equation (\ref{coef_perturb_modif}), the coefficients $h_{_\textrm{R}}$ and
$f_{_\textrm{R}}$ contain the general-relativistic correction of the
magnetic-field intensity at the stellar surface and are expressed
as\footnote{Note that $h_{_\textrm{R}} = 1/2$ and $f_{_\textrm{R}} = 1$ in the Newtonian
  limit of ${M/R\rightarrow 0}$}
\begin{equation}
\label{gr_fnctn}
h_{_\textrm{R}}=\frac{3 R^2 N_{_\textrm{R}}}{8 M^2}
    \left[\frac{R}{M}\ln N_{_\textrm{R}}^2 +\frac{1}{N_{_\textrm{R}}^2}+ 1
    \right] \,, \quad  \qquad \quad
f_{_\textrm{R}}=-\frac{3R^3}{8M^3}\left[\ln N_{_\textrm{R}}^2+\frac{2M}{R}
    \left(1+\frac{M}{R}\right)\right]\,,
\end{equation}
so that the unperturbed magnetic-field components at the stellar
surface can be written in a form that is reminiscent of the
Newtonian expressions, \ie
\begin{eqnarray}
\label{mf_1} && B_{_\textrm{R}}^{\hat r} = f_{_\textrm{R}} B\; (\cos\chi \cos\theta
+
    \sin\chi \sin\theta \,{\rm e}^{{\rm i}\phi})
    \,,
\\\nonumber\\
\label{mf_2} && B_\textrm{R}^{\hat \theta} = h_{_\textrm{R}}B\; (\cos\chi \sin\theta
    - \sin\chi \cos\theta \,{\rm e}^{{\rm i}\phi})
    \,,
\\\nonumber\\
\label{mf_3} && B_\textrm{R}^{\hat \phi} = -{\rm i} h_{_\textrm{R}}B\; \sin\chi
\,{\rm e}^{{\rm i}\phi}
    \,.
\end{eqnarray}

Two remarks are worth making at this point. First, as a consequence of
the assumption of infinite conductivity for the stellar material, the
magnetic field is advected with the fluid; this is expressed by the
'frozen-flux' condition for the perturbed magnetic field
\begin{equation}
\label{ff}
\partial_{\rm t} \delta \vec{\boldsymbol{B}} = \nabla \times
\left(\vec{\boldsymbol{v}} \times \vec{\boldsymbol{B}}\right) \,,
\end{equation}
and is satisfied by equations (\ref{nz_mfg1})--(\ref{nz_mfg3}). Secondly,
expressions (\ref{nz_mfg1})--(\ref{nz_mfg3}) refer to a generic velocity
field and thus lead to the integration constants
(\ref{coef_perturb_modif}) that are totally general.

\subsection{General expressions for the electric field}

The general expression for the electric fields can be obtained following
the same procedure adopted for the magnetic field in the previous
section. In particular, we recall that the solution for the vacuum
electric field in the near zone produced via a perturbation of a dipolar
magnetic field can be written as [see equations 78--80 of paper I]
\begin{eqnarray}
\label{nz_ef1}
&& \delta E^{\hat r} = \frac{M^2}{r^2}\ell^2\left(\ell+1\right)
    \left[Q_{\ell -1} -\left(1+\frac{r}{M}\ell \right)
    Q_{\ell} \right]
    \delta t_{\ell m}Y_{\ell m}\,,
\\ \nonumber\\
\label{nz_ef2}
&& \delta E^{\hat\theta} = \frac{M^2}{r^2N}\ell^2\left(\ell+1\right)
    \left[\left(1-\frac{r}{M}\right)
    Q_{\ell} - Q_{\ell -1}\right]
    \delta t_{\ell m}\,\partial_{\theta}Y_{\ell m}
    + \frac{r}{2N\sin\theta}\left[\ln N^2+\frac{2M}{r}
    \left(1+\frac{M}{r}\right)\right]
    \delta x_{1m} \,\partial_{\phi}Y_{1m} \,,
\\ \nonumber\\
\label{nz_ef3}
&& \delta E^{\hat\phi} = \frac{M^2}{r^2N\sin\theta}\ell^2\left(\ell+1\right)
    \left[\left(1-\frac{r}{M}\right)
    Q_{\ell} - Q_{\ell -1}\right]
    \delta t_{\ell m}\,\partial_{\phi}Y_{\ell m}
    - \frac{r}{2N}\left[\ln N^2+\frac{2M}{r}
    \left(1+\frac{M}{r}\right)\right]
    \delta x_{1m}\,\partial_{\theta} Y_{1m} \,,
\end{eqnarray}
where $Q_{\ell} = Q_{\ell}(x)= Q_{\ell}(1-r/M)$ is the Legendre function
of the second kind \citep{Arfken2005}. Note that, in general, the perturbed
electric field will have a zero background value but a non-vanishing
first-order perturbation which is proportional to the first-order
velocity perturbation and to the zero-th order magnetic field, \ie
$|\delta E| \propto |\delta v B_{_0}|$.

As in paper I, the integration constants $\delta t_{\ell m}$ and $\delta
x_{1m}$ can be calculated from imposing the continuity of the tangential
components of the electric field, and thus through expressions involving
the values of the magnetic field and of the velocities at the stellar
surface as
\begin{eqnarray}
\label{tlm}
&& \delta t_{\ell m}(t) = \frac{R^2}{\ell^3\left(\ell+1\right)^2M^2}
    \left[\left(1-\frac{R}{M}\right)
    Q_{\ell}(x_{_\textrm{R}}) - Q_{\ell -1}(x_{_\textrm{R}})\right]^{-1}
\nonumber \\
&& \hskip 5.0 cm
    \times \int {\rm d}\Omega\left\{\partial_{\theta}Y^*_{\ell m}
        \left[\delta v_{_\textrm{R}}^{\hat\phi}(t) B^{\hat r}_{_\textrm{R}}
        -\delta v_{_\textrm{R}}^{\hat r}(t) B^{\hat\phi}_{_\textrm{R}}\right]
        -{\rm i}\frac{mY^*_{\ell m}}{\sin\theta}
        \left[\delta v_{_\textrm{R}}^{\hat\theta}(t) B^{\hat r}_{_\textrm{R}}
        -\delta v_{_\textrm{R}}^{\hat r}(t) B^{\hat\theta}_{_\textrm{R}}\right]\right\} \,,
\\ \nonumber \\
\label{xlm}
&& \delta x_{1m}(t) = -\frac{3R^2}{8M^3f_{_\textrm{R}}}
    \int {\rm d}\Omega\left\{\partial_{\theta}Y^*_{1m}
        \left[\delta v_{_\textrm{R}}^{\hat\theta}(t) B^{\hat r}_{_\textrm{R}}
        -\delta v_{_\textrm{R}}^{\hat r}(t) B^{\hat\theta}_{_\textrm{R}}\right]
        +{\rm i}\frac{mY^*_{1m}}{\sin\theta}
        \left[\delta v_{_\textrm{R}}^{\hat\phi}(t) B^{\hat r}_{_\textrm{R}}
        -\delta v_{_\textrm{R}}^{\hat r}(t) B^{\hat\phi}_{_\textrm{R}}\right]\right\} \,,
\end{eqnarray}
where $Q_{\ell}(x_{_\textrm{R}}) := Q_{\ell}(1-R/M)$.

Another aspect of the interconnection between electric and magnetic
fields worth remarking is expressed in equations
(\ref{nz_ef1})--(\ref{nz_ef3}), where a magnetic field with multipolar
components up to $\ell$ induces a perturbed electric field with
multipolar components up to $(\ell + 1)$. As an aid for the calculations
which will be presented in the following, we report the explicit
expressions of the coefficients multiplying the integral in equation
(\ref{tlm}) for some relevant values of the multipoles
\begin{equation}
\frac{R^2}{\ell^3\left(\ell+1\right)^2M^2}
    \left[\left(1-\frac{R}{M}\right)
    Q_{\ell}(x_{_\textrm{R}}) - Q_{\ell -1}(x_{_\textrm{R}})\right]^{-1}=
\begin{cases}
{3R^3}/({16M^3N_{_\textrm{R}}h_{_\textrm{R}}})    & \text{for~ $\ell = 1$\,,}\\\\
{1}/({54N_{_\textrm{R}}^2g_{_\textrm{R}}})        & \text{for~ $\ell = 2$\,,}\\\\
{1}/({432 N_{_\textrm{R}}^2k_{_\textrm{R}}})      & \text{for~ $\ell = 3$\,,}\\\\
-{3M^3}/({100 R^3\gamma_{_\textrm{R}}}) & \text{for~ $\ell = 4$\,.}
\end{cases}
\end{equation}
Here $g_{_\textrm{R}}$ is a constant coefficient given by [see equation
  129 of paper I]
\begin{equation}
\label{g_R}
g_{_\textrm{R}}:=
    \left(1-\frac{R}{M}\right)
    \ln N^2_{_{_\textrm{R}}} - \frac{2 M^2}{3R^2 N^2_{_{_\textrm{R}}}} - 2 =
    \left(1-\frac{R}{M}\right)\ln \left(1-\frac{2M}{R}\right)
    - \frac{2}{3}\left(\frac{M}{R}\right)^2 \frac{R}{R-2M} - 2
    \,,
\end{equation}
while the functions $k_{_\textrm{R}}$ and $\gamma_{_\textrm{R}}$ have explicit expressions
\begin{eqnarray}
&&k_{_\textrm{R}} := \left\{-\frac{1}{3}\left(1-\frac{R}{M}\right)
\left(\frac{M^2}{R^2N_{_\textrm{R}}^2}+
\frac{15}{2}\right)+\left[1+\frac{5R^2N_{_\textrm{R}}^2}{4M^2}\right]\ln
N_{_\textrm{R}}^2\right\}\,,\\ \nonumber\\
&&
\gamma_{_\textrm{R}} := \frac{2M}{R}\left(105N_{_\textrm{R}}^4+95\frac{M^2}{R^2}N_{_\textrm{R}}^2+
6\frac{M^4}{R^4}\right)
+ N_{_\textrm{R}}^2\left(1-\frac{M}{R}\right)\left(7N_{_\textrm{R}}^2+4\right)\ln
N_{_\textrm{R}}^2\,.
\end{eqnarray}
Note that in the Newtonian limit, \ie when $M/R \rightarrow 0$, the
coefficients $g_{_\textrm{R}}, \gamma_{_\textrm{R}}$ and $k_{_\textrm{R}}$ go to zero, but the
integration constants converge to finite values.

\subsection{Main properties of spheroidal oscillations}
\label{spheroid}

While expressions (\ref{nz_mfg1})--(\ref{nz_mfg3}) and
(\ref{nz_ef1})--(\ref{nz_ef3}) are particularly effective because of
their completeness and generality, they are not particularly useful if
not specialized to a specific perturbation experienced by the star. In
view of this, in the following sections, we concentrate on the form that
these expressions attain when referred to the most common modes of
oscillation and using the mode nomenclature of
\citet{McDermott1988}. Before doing that, here, we briefly recall the
main properties of the typical oscillation modes of relativistic neutron
stars. Much of this material is well known but we recall it here for
completeness and because it will turn out useful in the subsequent
discussion. More information can be found in the review by
\citet{Kokkotas99a}.

When modelled as having a fluid core, a solid crust, and a thin surface
fluid 'ocean', neutron stars are capable of sustaining a broad variety
of normal modes of oscillation. For any star, there are two general
categories of non-radial oscillations: {spheroidal} (or polar) modes
and {toroidal} (or axial) modes.

Spheroidal modes, in particular, include several subclasses: the $p$ ,
$f$, and $g$ modes, which are well known from conventional stellar
pulsation theory, but also $s$ and $i$ modes, which result from crustal
elasticity and play a particularly important role in neutron stars. $p$
modes have pressure gradients as the main restoring force. The
eigenfunctions of a $p$ mode with quantum number $n$ has $n$ nodes and,
as $n$ increases, the frequency increases and the wavelength becomes
smaller. In the limit of short wavelengths, these modes represent simple
acoustic waves travelling in the star. In the same limit, the frequency of
the mode will tend to infinity. For a 'canonical' neutron star, \ie a
neutron star with mass $M \simeq 1.4\,M_\odot$ and radius $R \simeq 14\,
{\rm km}$, the typical frequency of the lowest order $p$-mode is a few
${\rm kHz}$ and are roughly proportional to the average rest-mass
density. $g$, on the other hand, modes have a buoyancy force (produced,
for instance, by gradients in temperature, composition or density) as the
main restoring force. There are two groups of $g$ modes: the 'core' $g$
modes, which are displacements confined almost completely to the fluid
core, and the 'surface' $g$ modes, which are limited primarily to the
thin fluid layer overlaying the crust. The core $g$ modes have typical
frequencies of $\sim 0.1 \,{\rm kHz}$, while the surface $g$ modes have a
typical frequency of $\sim 10 \, {\rm Hz}$. The $g$-mode frequencies are
roughly proportional to the internal temperature. $f$ modes have a
character which is intermediate between those of $p$- and $g$ modes and
are also referred to as the fundamental modes of oscillation. For each
$\ell$, the frequency of this mode is between the lowest order $g$-mode
(\ie the highest frequency $g$-mode) and the lowest order $p$-mode (\ie
the lowest frequency $p$-mode). Note that for a given pair of quantum
numbers ($\ell,m$), only one $f$-mode exists and its eigenfunctions have
no nodes. The typical frequency of the lowest order $f$-mode for a
canonical non-rotating neutron star is $\sim 2 - 3 \,{\rm kHz}$. $s$ modes
are essentially normal modes of shear waves in the solid neutron star
crust. These modes have quadrupole frequencies $\sim {\rm kHz}$ and
depend strongly on the crust thickness. Waves can also propagate on the
solid-fluid (crust-core) interface, and the normal modes corresponding to
such waves are the interface, or $i$ modes; these modes resemble
acoustic waves scattered off a hard sphere and do not induce significant
fluid motion. The frequencies of these modes depend strongly on the local
density and temperature at the interface, but are normally with
frequencies of a few kHz or higher.

Finally, toroidal modes are modes that, unsurprisingly, have
eigenfunctions described by purely toroidal (axial) functions and hence
do not have radial displacements. $r$ modes, in particular, are a well-known
member of this class of modes and have the Coriolis force as
restoring force; in this respect, however, $r$ modes represent more an
exception than a rule. Toroidal modes (and hence $r$ modes) are, in fact,
part of a larger class of modes having the Coriolis force as the main
restoring force. Such modes are called {inertial} modes and have
eigenfunctions with a mixed spheroidal and toroidal nature, approximately
of the same magnitude, at least to first order in the slow-rotation
expansion. In this respect, $r$ modes can be seen as inertial modes with
purely toroidal eigenfunctions. The velocity eigenfunctions in the case
of toroidal modes have very simple expressions and the perturbations in
the density and pressure appear at orders higher than the first one in
the slow-rotation approximation.

We can now start our specialization of the general expressions presented
above by first considering the case of spheroidal oscillations, where the
Euler velocity field is given by [see, for example, equation 13.60 of
  \citet{Unno1989}]
\begin{equation}
\label{spheroidal_vf}
\delta v^{\hat i}=\left(\eta (r)
Y_{\ell^{\prime} m^{\prime}}
    (\theta ,\phi) \,,
    \xi (r)\,\partial_{\rm \theta}
    Y_{\ell^{\prime} m^{\prime}}(\theta ,\phi) \,,
    \frac{\xi (r)}{\sin\theta}\,\partial_{\rm \phi}
    Y_{\ell^{\prime} m^{\prime}}(\theta ,\phi)
    \right)\,{\rm e}^{-{\rm i}\omega t} \,,
\end{equation}
with $\omega$ being the real part of the oscillation frequency, while
$\eta(r)$ and $\xi(r)$ are the radial eigenfunctions. We next assume that
the oscillations are zero at the centre of the star, but non-vanishing at
its surface, \ie $\eta(0)=0$, and $\eta_{_\textrm{R}} := \eta(R)\ne 0$. Note that
in principle the eigenfunctions $\eta(r)$ and $\xi(r)$ should be
calculated through the solution of the corresponding eigenvalue problem.
Here, however, we will consider them as given functions, whose value at
the stellar surface will be used to estimate the electromagnetic
emission. Note also that we have used multipolar indices $\ell^{\prime}$
and $m^{\prime}$ to distinguish the harmonic dependence of the velocity
perturbations (\ref{spheroidal_vf}) from the harmonic dependence of the
electromagnetic fields, which we express in terms of the indices $\ell$
and $m$.

Since the stellar surface itself is undergoing oscillations also in the
radial direction, its radial position in time will be expressed as
\begin{equation}
\label{crust_position} R(t) \cong R_0 + \int_0^t \delta v^{\hat
r}(R_0,t') dt' \,,
\end{equation}
where $R_0$ is the position of the stellar surface at $t=0$. We next
assume that $\tilde{\xi}=\tilde{\xi}_0\,{\rm e}^{-{\rm i}\omega t}$ is
the \textit{transverse} displacement, and
$\tilde{\eta}=\tilde{\eta}_0{\rm e}^{-{\rm i}\omega t}$ is the
displacement of the stellar crust in the {radial} direction, with
$\dot{\tilde{\eta}}=-{\rm i}\omega\tilde{\eta}\ll 1$. From the condition
of quasi-stationarity, \ie $|\delta v|/c=\omega\tilde{\eta}/c\ll 1$, one
can estimate that the normalized radial displacement $\epsilon
={\tilde{\eta}}/{R_0}\ll 1$ for typical oscillations in the kHz range. As
a result, the radial position of the stellar surface
(\ref{crust_position}) is
\begin{equation}
\label{crust_position_1} R(t) = R_0 + \frac{\tilde{\eta}}{R_0}R_0=
R_0(1+\epsilon(t)) \,.
\end{equation}

In principle, the dipolar magnetic field at the stellar surface will also
change a result of the changes in the position of the stellar surface and
assume a general form of type
\begin{equation}
B (t,R)=\frac{2\mu(t)}{R^3(t)}\simeq \frac{2\mu_0}{R^3(t)} \simeq
\frac{2\mu_0}{R_{_0}^3}\left[1-3\epsilon(t)\right]=
{B}_{_0} (1-3\epsilon(t)) \,,
\end{equation}
where ${B}_{_0}:= 2\mu/R_{_0}^3$. However, the time derivative of the
dipolar magnetic field at the dynamical boundary of the star is then
\begin{equation}
\partial_{\rm t} B (t,R)= -3 {B}_{_0} \,\partial_{\rm t} \epsilon =
    3 {\rm i}\omega \frac{\eta}{R_{_0}}{B}_{_0}= \mathcal{O}(\epsilon^2) \,,
\end{equation}
since the terms proportional to $\epsilon$ appear as product
$\epsilon\omega R_{_0}$ and are negligible in the linear approximation
where $\epsilon\ll 1$ and $\omega R_{_0}\ll 1$.

Consider now a relativistic star having a background dipolar magnetic
field (\ie with $\ell=1$) not necessarily aligned with the polar axis
(\ie with $\chi \ne 0$) and undergoing oscillations with components
(\ref{spheroidal_vf}). The integration constants
(\ref{coef_perturb_modif}) for the spheroidal oscillations of the dipolar
magnetic field then take the form
\begin{eqnarray}
\label{coef_perturb_spheroid}
&& \partial_{\rm t} \delta s_{1m}(t)=-{\rm i}\omega_{_\textrm{R}}\delta s_{1m} =
    -\frac{3R^2}{8f_{_\textrm{R}}}B_{_0} \,{\rm e}^{-{\rm i}\omega_{_\textrm{R}} t}
    \int {\rm d}\Omega Y^*_{1 m}\Bigg\{
    \left[2h_{_\textrm{R}}\eta_{_\textrm{R}}+f_{_\textrm{R}}\xi_{_\textrm{R}}\ell'(\ell'+1)\right]
    Y_{\ell'm'}
    \left(\cos\theta\cos\chi+\sin\theta\sin\chi \,{\rm e}^{{\rm i}\phi}\right)
\nonumber\\\nonumber\\
&&
\hskip 3.75 cm
    +\left(h_{_\textrm{R}}\eta_{_\textrm{R}}+f_{_\textrm{R}}\xi_{_\textrm{R}}\right)
    \left[Y_{\ell'm',\theta}
    \left(\sin\theta\cos\chi-\cos\theta\sin\chi \,{\rm e}^{{\rm i}\phi}\right)
    -\frac{{\rm i}}{\sin\theta}\sin\chi \,{\rm e}^{{\rm i}\phi} Y_{\ell'm',\phi}\right]
    \Bigg\}\,.
\end{eqnarray}
When $\ell'=2, m'=1$, the integration constants
(\ref{coef_perturb_spheroid}) take the form
\begin{eqnarray}
&&\delta s_{10}= 0 \,,
\\ \nonumber\\
&& \delta s_{11}= \rm{i}\frac{3}{8\sqrt{5}}
    \frac{R^2}{f_{_\textrm{R}}\omega_{_\textrm{R}}} B_{_0}
    \left(h_{_\textrm{R}}\eta_{_\textrm{R}}-3f_{_\textrm{R}}\xi_{_\textrm{R}}\right)
        \,{\rm e}^{-{\rm i}\omega_{_\textrm{R}}t} \cos\chi\,.
\end{eqnarray}
As a result, the components of the perturbed magnetic field
(\ref{nz_mfg1})--(\ref{nz_mfg3}) in the near zone are the real
parts of the following complex expressions
\begin{eqnarray}
\label{nz_mf_spheroid1}
&& \delta B^{\hat r} = -{\rm
i}\sqrt{\frac{3}{10\rm{\pi}}}
    \frac{3R^2}{16M^3f_{_\textrm{R}}\omega_{_\textrm{R}}}
    \left[\ln N^2+
    \frac{2M}{r}\left(1+\frac{M}{r}\right)\right]
    B_{_0}
    \left(h_{_\textrm{R}}\eta_{_\textrm{R}}-3f_{_\textrm{R}}\xi_{_\textrm{R}}\right)
        \,{\rm e}^{-{\rm i}(\omega_{_\textrm{R}}t-\phi)}
    \cos\chi\sin\theta
\,,
\\ \nonumber \\
\label{nz_mf_spheroid2}
&& \delta B^{\hat \theta} = -{\rm
i}\sqrt{\frac{3}{10\rm{\pi}}}
    \frac{3R^2N}{16M^2rf_{_\textrm{R}}\omega_{_\textrm{R}}}
    \left(\frac{r}{M}\ln N^2+
        \frac{1}{N^2}+1\right)
    B_{_0}
    \left({h_{_\textrm{R}}\eta_{_\textrm{R}}}-3{f_{_\textrm{R}}}\xi_{_\textrm{R}}\right)
        \,{\rm e}^{-{\rm i}(\omega_{_\textrm{R}}t-\phi)}
    \cos\chi\cos\theta
\,,
\\ \nonumber \\
\label{nz_mf_spheroid3}
 && \delta B^{\hat \phi} =
\sqrt{\frac{3}{10 \rm{\pi}}}
    \frac{3R^2N}{16M^2rf_{_\textrm{R}}\omega_{_\textrm{R}}}
    \left(\frac{r}{M}\ln N^2+\frac{1}{N^2}+1\right)
    B_{_0}
    \left({h_{_\textrm{R}}\eta_{_\textrm{R}}}-3{f_{_\textrm{R}}}\xi_{_\textrm{R}}\right)
        \,{\rm e}^{-{\rm i}(\omega_{_\textrm{R}}t-\phi)} \cos\chi
    \,,
\end{eqnarray}
where we recall that $\omega_{_\textrm{R}}$ is the angular frequency of
oscillation measured by an observer at the stellar surface.

The oscillation frequency, in fact, is itself subject to the standard
gravitational redshift, so that the frequency $\omega(r)$ at a generic
position $r > R$ will be redshifted and given by [see equation 87 of
  paper I]
\begin{equation}
\label{redshift}
\omega(r) = \omega_{_\textrm{R}} \frac{N_{_\textrm{R}}}{N} =
    \omega_{_\textrm{R}} \sqrt{\left(\frac{R-2M}{r-2M}\right)\frac{r}{R}} \,,
\end{equation}
and asymptoting to $\omega = \omega_{_\textrm{R}} \sqrt{1-2M/R}$ at spatial
infinity. For clarity, hereafter we will indicate as $\omega(r)$ the
function given by expression \eqref{redshift} and as $\omega$ the value
$\omega_{_\textrm{R}} \sqrt{1-2M/R}$, when $r \to \infty$.

For completeness, we also report the Newtonian limits of the
time-dependent part of the perturbed magnetic field in the near zone
\begin{eqnarray}
\label{nz_mf_spheroid1_flat}
 && \delta B^{\hat r} = {\rm
i}\sqrt{\frac{3}{10 \rm{\pi}}}
    \frac{1}{4\omega R}
                  \left(\frac{R}{r}\right)^3 B_{_0}
    \left(\eta_{_\textrm{R}}-6\xi_{_\textrm{R}}\right)
        \,{\rm e}^{-{\rm i}(\omega t -\phi)}
    \cos\chi\sin\theta
 = - {2}{\tan \theta} \,\delta B^{\hat \theta} = 2{\rm i}
    \sin\theta \,\delta B^{\hat \phi} \,,
\end{eqnarray}
corresponding to the Newtonian limit of equations
(\ref{nz_mf_spheroid1})--(\ref{nz_mf_spheroid3}) and not reported in the
work of \citet{Muslimov1986}.

Proceeding in a similar manner, the calculation of the near-zone electric
field for oscillation modes with $\ell'=2, m'=1$ gives the following
non-vanishing integration constants
\begin{equation}
\label{tlm_spher_nz_10}
\delta t_{21}=\frac{\rm i}{54}
    \frac{h_{_\textrm{R}}}{N^2_{_\textrm{R}}g_{_\textrm{R}}}B_{_0}\eta_{_\textrm{R}}
     \cos\chi \,{\rm e}^{-{\rm i}\omega_{_\textrm{R}}t} \,, \qquad
\delta x_{11}=
    \frac{3R^2}{8\sqrt{5}f_{_\textrm{R}}M^3}B_{_0}
    \left({h_{_\textrm{R}}}\eta_{_\textrm{R}}+2f_{_\textrm{R}}\xi_{_\textrm{R}}\right)
        \,{\rm e}^{-{\rm i}\omega_{_\textrm{R}}t} \cos\chi\,,
\end{equation}
and, consequently, the following components of the near-zone electric
fields
\begin{eqnarray}
\label{nz_ef1_spheroid_21}
&&\delta E^{\hat r}=-{\rm i}\sqrt{\frac{5}{6\rm{\pi}}}
    \frac{h_{_\textrm{R}}}{4 g_{_\textrm{R}} N_{_{_\textrm{R}}}}
    \left[\left(3-\frac{2r}{M}\right)
    \ln N^2 + \frac{2M^2}{3r^2}+\frac{2M}{r}-4\right]
    B_{_0}\eta_{_\textrm{R}}\,{\rm e}^{-{\rm i}(\omega_{_\textrm{R}}t-\phi)}
\cos\chi\cos\theta\sin\theta \,,
\\ \nonumber\\
&&\label{nz_ef2_spheroid_21} \delta E^{\hat\theta}= -\frac{{\rm
i}}{4}\Bigg\{\sqrt{\frac{5}{6\rm{\pi}}}
    \frac{N}{g_{_\textrm{R}} N_{_{_\textrm{R}}}^2}
 \left[\left(1-\frac{r}{M}
   \right)\ln N^2-2-\frac{2M^2}{3r^2N^2}\right]
 {h_{_\textrm{R}}}\eta_{_\textrm{R}}
    \cos 2\theta  + \nonumber
\\ \nonumber \\
    && \hskip 2.5cm\sqrt{\frac{3}{10\rm{\pi}}}\frac{3rR^2}{8M^3Nf_{_\textrm{R}}}
    \left[\ln N^2+
    \frac{2M}{r}\left(1+\frac{M}{r}\right)\right]
         \left({h_{_\textrm{R}}}\eta_{_\textrm{R}}+2f_{_\textrm{R}}\xi_{_\textrm{R}}\right)
         \Bigg\}B_{_0}
        \,{\rm e}^{-{\rm i}(\omega_{_\textrm{R}}t-\phi)}\cos\chi \,,
\\ \nonumber\\
&&\label{nz_ef3_spheroid_21} \delta E^{\hat\phi}= \frac{{\rm
1}}{4}\Bigg\{\sqrt{\frac{5}{6\rm{\pi}}}
    \frac{N}{g_{_\textrm{R}} N_{_{_\textrm{R}}}^2}
 \left[\left(1-\frac{r}{M}
   \right)\ln N^2-2-\frac{2M^2}{3r^2N^2}\right]
 {h_{_\textrm{R}}}\eta_{_\textrm{R}}  + \nonumber
\\ \nonumber \\
    && \hskip 2.5cm\sqrt{\frac{3}{10\rm{\pi}}}\frac{3rR^2}{8M^3Nf_{_\textrm{R}}}
    \left[\ln N^2+
    \frac{2M}{r}\left(1+\frac{M}{r}\right)\right]
         \left({h_{_\textrm{R}}}\eta_{_\textrm{R}}+2f_{_\textrm{R}}\xi_{_\textrm{R}}\right)
         \Bigg\}B_{_0}
        \,{\rm e}^{-{\rm i}(\omega_{_\textrm{R}}t-\phi)}\cos\chi\cos\theta \,.
\end{eqnarray}
Finally, we note that the Newtonian limits of the expressions for the
electric fields (\ref{nz_ef1_spheroid_21}) and (\ref{nz_ef3_spheroid_21})
reduce to the Newtonian solutions found by Muslimov \& Tsygan [\cf
  equation 19 with the integration constants defined by expressions A.1
  and A.2 of \citet{Muslimov1986}].

\subsubsection{Radial oscillations}
\label{ro}

Next, in analogy with what done by \citet{McDermott1988}, we consider the
special case of spheroidal oscillations that are purely radial and thus a
velocity field given by
\begin{equation}
\label{rad_puls}
\delta v^{\hat i}(r,t):=
\biggl(\eta(r),0,0\biggr)\,{\rm e}^{-{\rm i}\omega t}\,,
\end{equation}
where $\eta(r)$ is the eigenfunction. Using equations
(\ref{nz_mfg1})--(\ref{nz_mfg3}) with the condition for the integration
constant $\delta s_{\ell m}$ expressed as
\begin{eqnarray}
\label{coef_perturb_radial_10}
&&\delta s_{10}(t)=-{\rm i}\sqrt{\frac{3\rm{\pi}}{4}}
    \frac{h_{_\textrm{R}}R^2}{f_{_\textrm{R}}\omega_{_\textrm{R}}}B_{_0}
    \eta_{_\textrm{R}}\,{\rm e}^{-{\rm i}\omega_{_\textrm{R}}t}\cos\chi
=-\frac{1}{\sqrt{2}\tan \chi} \delta s_{11}(t) \,,
\end{eqnarray}
the perturbed magnetic field will have components given by the
expressions
\begin{eqnarray}
\label{nz_pert_mf_r}
&& \delta B^{\hat r} = -{\rm
i}\frac{3R^2h_{_\textrm{R}}}{4M^3f_{_\textrm{R}}\omega_{_\textrm{R}}}
        \left[\ln N^2 +
    \frac{2M}{r} \left(1 +  \frac{M}{r}\right) \right]
        B_{_0}\eta_{_\textrm{R}}\,{\rm e}^{-{\rm i}\omega_{_\textrm{R}}t}
    \left(\cos\theta\cos\chi + \sin\theta\sin\chi \,{\rm e}^{{\rm i}\phi}\right)\,,
\\\nonumber \\
\label{nz_pert_mf_theta}
&& \delta B^{\hat \theta} =  {\rm
i}\frac{3R^2Nh_{_\textrm{R}}}{4M^2rf_{_\textrm{R}}\omega_{_\textrm{R}}}
        \left(\frac{r}{M}\ln N^2+
    \frac{1}{N^2}+1\right)
         B_{_0}\eta_{_\textrm{R}}\,{\rm e}^{-{\rm i}\omega_{_\textrm{R}}t}
    \left(\sin\theta\cos\chi - \cos\theta\sin\chi \,{\rm e}^{{\rm i}\phi}\right)\,,
\\\nonumber \\
\label{nz_pert_mf_phi}
&& \delta B^{\hat \phi} =
\frac{3R^2Nh_{_\textrm{R}}}{4M^2rf_{_\textrm{R}}\omega_{_\textrm{R}}}
        \left(\frac{r}{M}\ln N^2+
    \frac{1}{N^2}+1\right)
         B_{_0}\eta_{_\textrm{R}}\,{\rm e}^{-{\rm i}\omega_{_\textrm{R}}t}\sin\chi \,{\rm e}^{{\rm i}\phi} \,.
\end{eqnarray}
It should be noted that the integration constants
(\ref{coef_perturb_radial_10}) could be easily obtained from the
integration constants for spheroidal modes (\ref{coef_perturb_spheroid})
if one replaced  $Y_{\ell'm'}$ everywhere with $1$; this is simply because
in such a case, the spheroidal velocity field (\ref{spheroidal_vf})
coincides with the radial one (\ref{rad_puls}). It is also easy to realize,
after using the expansion of the magnetic field 56--57 of paper I,
the radial velocity field (\ref{rad_puls}) in expression (\ref{tlm}) and
the condition $\partial_{\phi}Y_{\ell m}={\rm i}mY_{\ell m}$, that the
integration constants $t_{\ell m}$ are identically zero since
\begin{equation}
-\int {\rm d}\Omega\left(\partial_{\theta}Y^*_{\ell m}
        \,\delta v_{_{_\textrm{R}}}^{\hat r}B^{\hat\phi}_{_{_\textrm{R}}}
        -{\rm i}\frac{mY^*_{\ell m}}{\sin\theta}
        \,\delta v_{_{_\textrm{R}}}^{\hat r}B^{\hat\theta}_{_{_\textrm{R}}}\right)=
    -\delta v_{_\textrm{R}}^{\hat r}\frac{N_{_\textrm{R}}}{R}\,\partial_{\textrm{r}}S_{\ell
    m}|_{_{r=R}}
    \int {\rm d}\Omega\left(\partial_{\theta}Y^*_{\ell m}
        \frac{1}{\sin\theta}\,\partial_{\phi}Y_{\ell m}
        -{\rm i}\frac{mY^*_{\ell m}}{\sin\theta}
        \,\partial_{\theta}Y_{\ell m}\right)=0 \,.
\end{equation}
As a result, the only non-vanishing integration constant is
\begin{eqnarray}
&& \delta x_{10}= -\frac{\sqrt{3\rm{\pi}}}{2}
    \frac{R^2}{f_{_\textrm{R}}M^3}B_{_0}
    h_{_\textrm{R}}\eta_{_\textrm{R}}
        \,{\rm e}^{-{\rm i}\omega_{_\textrm{R}}t} \cos\chi\ =
        \frac{\sqrt{2}}{\tan \chi}\delta x_{11} \,,
\end{eqnarray}
and the corresponding electric field has non-vanishing components
\begin{eqnarray}
\label{e_rad_puls_2}
&&\delta E^{\hat\theta}=
    -\frac{3 rR^2}{16M^3N}\frac{h_{_\textrm{R}}}{f_{_\textrm{R}}}
    \left[\ln N^2+
    \frac{2M}{r}\left(1+\frac{M}{r}\right)\right]
        B_{_0} \eta_{_\textrm{R}}\,{\rm e}^{-{\rm i}\omega_{_\textrm{R}} t}\sin\chi\sin\phi \,,
\\ \nonumber\\
\label{e_rad_puls_3}
&&\delta E^{\hat\phi}=
    \frac{3 rR^2}{16M^3N}\frac{h_{_\textrm{R}}}{f_{_\textrm{R}}}
    \left[\ln N^2+
    \frac{2M}{r}\left(1+\frac{M}{r}\right)\right]
        B_{_0}\eta_{_\textrm{R}}\,{\rm e}^{-{\rm i}\omega_{_\textrm{R}} t}
    \left(2\sin\theta\cos\chi+
    \sin\chi\cos\theta\cos\phi\right) \,.
\end{eqnarray}

It is also useful to point out that the solutions
(\ref{e_rad_puls_2}) and (\ref{e_rad_puls_3}) guarantee the force-free
condition $E^{\hat i}B_{\hat i}=0$, thus expressing the fact that no
particle acceleration is possible along the perturbed electromagnetic
fields within the near zone. Furthermore, expressions
(\ref{e_rad_puls_2}) and (\ref{e_rad_puls_3}) have a rather simple
interpretation. Consider in fact a magnetic dipole that is aligned with
the polar axis (\ie with $\chi=0$); in this case, the magnetic field will
then be only poloidal and any perturbation of this magnetic field will
produce a new, oscillating electric field that will be purely toroidal
(\ie $\delta E^{\hat\theta}=0$).

Also in this case, the Newtonian limits of the expressions for the
electric fields (\ref{e_rad_puls_2}) and (\ref{e_rad_puls_3}) reduce to
the Newtonian solutions of \citet{Muslimov1986} [\cf equation B2 of
  Muslimov \& Tsygan (1986)], while the corresponding Newtonian limits of
the near-zone magnetic fields
(\ref{nz_pert_mf_r})--(\ref{nz_pert_mf_phi}) are given by
\begin{eqnarray}
\label{nz_pert_mf_r_flat}
&& \delta B^{\hat r} = -{\rm
i}\left(\frac{R}{r}\right)^3\frac{1}{\omega R}
        B_{_0}\eta_{_\textrm{R}}\,{\rm e}^{-{\rm i}\omega t}
    \left(\cos\theta\cos\chi + \sin\theta\sin\chi \,{\rm e}^{{\rm i}\phi}\right)\,,
\\\nonumber \\
\label{nz_pert_mf_theta_flat}
&& \delta B^{\hat \theta} = {\rm
i}\left(\frac{R}{r}\right)^3\frac{1}{2\omega R}
        B_{_0}\eta_{_\textrm{R}}\,{\rm e}^{-{\rm i}\omega t}
    \left(\sin\theta\cos\chi - \cos\theta\sin\chi \,{\rm e}^{{\rm i}\phi}\right)\,,
\\\nonumber \\
\label{nz_pert_mf_phi_flat}
&& \delta B^{\hat \phi} =
\left(\frac{R}{r}\right)^3\frac{1}{2\omega R}
        B_{_0}\eta_{_\textrm{R}}\,{\rm e}^{-{\rm i}\omega t}
         \sin\chi \,{\rm e}^{{\rm i}\phi} \,.
\end{eqnarray}
\subsection{Toroidal oscillations}
\label{to}

We next examine a more complex velocity field and, in particular, assess
the impact that toroidal oscillations may have on the electromagnetic
fields of the relativistic star. To this scope we consider a perturbative
velocity field with components [\cf equation 13.71 of \citet{Unno1989}]
\begin{equation}
\label{toroidal_vf}
\delta v^{\hat i}=\left(0,
    \frac{1}{\sin\theta}\,\partial_{\rm \phi}
    Y_{\ell^{\prime} m^{\prime}}(\theta ,\phi) \,,
    -\partial_{\rm \theta}
    Y_{\ell^{\prime} m^{\prime}}(\theta ,\phi)
    \right)\eta (r)\,{\rm e}^{-{\rm i}\omega t} \,.
\end{equation}
The toroidal velocity field (\ref{toroidal_vf}) has an interest of its
own and has attracted considerable attention since it corresponds to the
one for $r$-mode oscillations when observed in the frame corotating with
the star. In this case, in fact, it has been shown that such modes may
lead to unstable oscillations \citep{Andersson1998, Friedman1998},
although such an instability in a newly born neutron star has shown to be
contrasted by the growth of differential rotation \citep{Rezzolla00,
  Sa2004, Friedman2015} and by the amplification of magnetic fields
\citep{Rezzolla2001, Rezzolla2001b, Sa2006} [see also
  \citet{Cuofano2010,Cuofano2012} for the extension to low-mass X-ray
  binaries].

We start by computing the integration constants
(\ref{coef_perturb_modif}) for the toroidal oscillations, which take the
form
\begin{equation}
\label{coef_perturb_toroid}  \partial_{\rm t} \delta s_{1m}(t)=-{\rm
i}\omega_{_\textrm{R}}\delta s_{1m} =
    -\frac{3R^2}{8}B_{_0}\eta_{_\textrm{R}} \,{\rm e}^{-{\rm i}\omega_{_\textrm{R}}t}
    \int {\rm d}\Omega Y^*_{1m}\Bigg\{
    \left(\sin\theta\cos\chi-\cos\theta\sin\chi \,{\rm e}^{{\rm i}\phi}\right)
    \frac{1}{\sin\theta}\,\partial_{\rm \phi} Y_{\ell'm'}
    +{\rm i}\sin\chi \,{\rm e}^{{\rm i}\phi} \,\partial_{\rm \theta} Y_{\ell'm'}
    \Bigg\}\,.
\end{equation}
The components of the perturbed magnetic field generated by the toroidal
oscillations with $\ell'=m'=1$ and when $\chi \neq 0$ are then given by
\begin{eqnarray}
\label{nz_pert_tor_mf_r}
&& \delta B^{\hat r} = -\sqrt{\frac{3}{8\rm{\pi}}}
        \frac{3R^2}{8M^3\omega_{_\textrm{R}}}\left[\ln N^2 +
    \frac{2M}{r} \left(1 +  \frac{M}{r}\right) \right]
        B_{_0}\eta_{_\textrm{R}} \,{\rm e}^{-{\rm i}(\omega_{_\textrm{R}}t-\phi)}
        \cos\chi\sin\theta
\,,
\\\nonumber \\
\label{nz_pert_tor_mf_phi}
&& \delta B^{\hat \theta} =
-\sqrt{\frac{3}{8\rm{\pi}}}
        \frac{3NR^2}{8M^2r\omega_{_\textrm{R}}}
        \left[\frac{r}{M}\ln N^2+
    \frac{1}{N^2}+1\right]B_{_0}
        \eta_{_\textrm{R}}\,{\rm e}^{-{\rm i}(\omega_{_\textrm{R}}t-\phi)}
        \cos\chi\cos\theta = {\rm i}\,\delta B^{\hat \phi}\cos\theta \,,
\end{eqnarray}
where we have used the non-vanishing integration constant
\begin{equation}
\delta s_{11}(t)=
    \frac{3R^2}{8\omega_{_\textrm{R}}}B_{_0}\eta_{_\textrm{R}}\cos\chi
    \,{\rm e}^{-{\rm i}(\omega_{_\textrm{R}}t-\phi)}\,.
\end{equation}
The Newtonian limit of these near-zone magnetic fields
(\ref{nz_pert_tor_mf_r})--(\ref{nz_pert_tor_mf_phi}) will take a form
\begin{eqnarray}
\label{nz_pert_tor_mf_r_flat} && \delta B^{\hat r} =
\sqrt{\frac{3}{8\rm{\pi}}}
        \frac{1}{\omega R}\left(\frac{R}{r}\right)^3
        B_{_0}\eta_{_\textrm{R}} \,{\rm e}^{-{\rm i}(\omega t-\phi)}
        \cos\chi\sin\theta=
        -2 \tan\theta \,\delta B^{\hat \theta} =
        2{\rm i} \sin\theta \,\delta B^{\hat \phi}
\,.
\end{eqnarray}

Similarly, the electric fields outside the oscillating magnetized star
for velocity oscillation modes with $\ell^{\prime}=1, m^{\prime}=0$ are
computed to be
\begin{eqnarray}
\label{nz_tor_ef1}
&&\delta E^{\hat r} = -\sqrt{\frac{3}{4\rm{\pi}}}
    \frac{f_{_\textrm{R}}}{3 g_{_\textrm{R}} N_{_{_\textrm{R}}}^2}
    \left[\left(3-\frac{2r}{M}\right)
    \ln N^2 + \frac{2M^2}{3r^2}+\frac{2M}{r}-4\right]
    B_{_0}\eta_{_\textrm{R}}
    \left(3 \cos^2\theta-1\right)\cos\chi
    \,{\rm e}^{-{\rm i}\omega_{_\textrm{R}}t} \,,
\\\nonumber \\
\label{nz_tor_ef2}
&&\delta E^{\hat \theta} = \sqrt{\frac{3}{4\rm{\pi}}}
    \frac{Nf_{_\textrm{R}}}{g_{_\textrm{R}} N_{_{_\textrm{R}}}^2}
 \left[\left(1-\frac{r}{M}
   \right)\ln N^2-2-\frac{2M^2}{3r^2N^2}\right]
   B_{_0}\eta_{_\textrm{R}}
    \left(\cos\chi\sin\theta\cos\theta\right)
    \,{\rm e}^{-{\rm i}\omega_{_\textrm{R}}t}  \,,
\end{eqnarray}
where we have used the following expression for the only non-vanishing
integration constant for these modes
\begin{equation}
\label{tlm_tor_nz_10}
\delta t_{20}=-\frac{1}{9\sqrt{15}}
    \frac{f_{_\textrm{R}}}{N^2_{_\textrm{R}}g_{_\textrm{R}}}B_{_0}\eta_{_\textrm{R}}
     \cos\chi \,{\rm e}^{-{\rm i}\omega_{_\textrm{R}}t} \,.
\end{equation}

Other integration constants for modes higher than $\ell^{\prime}=1$ are
\begin{equation}
\label{tlm_tor_nz_21} \delta t_{31}=\frac{3}{432\sqrt{70}}
    \frac{f_{_\textrm{R}}}{N^2_{_\textrm{R}}k_{_\textrm{R}}}B_{_0}\eta_{_\textrm{R}}
     \cos\chi \,{\rm e}^{-{\rm i}\omega_{_\textrm{R}}t} \,, \qquad
\delta t_{11}=\frac{3}{8\sqrt{5}}
    \frac{f_{_\textrm{R}}R^3}{N_{_\textrm{R}}h_{_\textrm{R}}M^3}B_{_0}\eta_{_\textrm{R}}
    \cos\chi \,{\rm e}^{-{\rm i}\omega_{_\textrm{R}}t}\,,
\qquad \ell^{\prime}=2, m^{\prime}=1\,,
\end{equation}
and
\begin{equation}
\label{tlm_tor_nz_32} \delta
t_{42}=-\frac{3}{25}\sqrt{\frac{3}{7}}
    \frac{M^3f_{_\textrm{R}}}{R^3\gamma_{_\textrm{R}}}B_{_0}\eta_{_\textrm{R}}
     \cos\chi \,{\rm e}^{-{\rm i}\omega_{_\textrm{R}}t} \,, \qquad
\delta t_{22}=\frac{1}{9\sqrt{7}}
    \frac{f_{_\textrm{R}}}{N_{_\textrm{R}}^2g_{_\textrm{R}}}B_{_0}\eta_{_\textrm{R}}
    \cos\chi \,{\rm e}^{-{\rm i}\omega_{_\textrm{R}}t}\,, \qquad
\ell^{\prime}=3, m^{\prime}=2 \,.
\end{equation}
Also in this case, the Newtonian limit for the electric fields
(\ref{nz_tor_ef1})--(\ref{nz_tor_ef2}) coincide with those reported by
\citet{Muslimov1986} [\cf equation (19) with the integration constants
  given by expressions (A.5) and (A.6) of \citet{Muslimov1986}].

\section{Electromagnetic fields in the wave-zone}
\label{em_wz}

In what follows we extend the work of
the previous sections by providing
the expressions for the
electromagnetic fields in the wave zone and that
will be employed when computing the expressions for the electromagnetic
energy losses. We first recall that this it is possible to obtain
wave-zone expressions in terms of the spherical Hankel functions, which
have a simple radial fall-off in the case of small arguments, \ie
\begin{equation}
\label{small} H_{\ell}(\omega r)\approx -{\rm i}(2\ell-1)!!
    (\omega r)^{-\ell-1}\,, \qquad
    DH_{\ell}(\omega r)\approx {\rm i}\ell(2\ell-1)!!
    (\omega r)^{-\ell-2}\omega
    = - \ell \frac{H_{\ell}}{r}\,,
    \qquad {\rm for} \qquad
    \omega r\approx   \omega_{_\textrm{R}}R\ll 1 \,,
\end{equation}
where $DH_{\ell}(\omega r) := r^{-1} \,\partial_{r} \left[rH_{\ell}
  (\omega r)\right]$. On the other hand, the Hankel functions exhibit a
typical oscillatory behaviour (in space) in the limit of large arguments
\citep{Arfken2005}, \ie
\begin{equation}
\label{large}
    H_{\ell}(\omega r)\approx (-{\rm i})^{\ell+1}
    \frac{\,{\rm e}^{{\rm i}\omega r}}{\omega r}\,,
    \qquad
    DH_{\ell}(\omega r)
    \approx (-{\rm i})^{\ell}\frac{\,{\rm e}^{{\rm i}\omega r}}{r}
    = {\rm i} \omega H_{\ell}\,,
    \qquad {\rm for} \qquad
    \omega r\rightarrow \infty \,.
\end{equation}

After some rather lengthy algebra, the components of the magnetic fields
in the far zone are found to be given by the general expressions [see
  equations 97--99 of paper I]
\begin{eqnarray}
\label{sol_wz_mf1}
&& B^{\hat r}  =  \frac{\,{\rm e}^{-{\rm i}\omega t}
    \sqrt{\ell\left(\ell+1\right)}}{r}
    H_{\ell}(\omega r)u_{\ell m}Y_{\ell m}\,,
\\\nonumber\\
\label{sol_wz_mf2}
&& B^{\hat\theta}  =  \frac{\,{\rm e}^{-{\rm i}\omega t}}
    {\sqrt{\ell(\ell+1)}}
    \left(DH_{\ell}(\omega r)u_{\ell m}\,\partial_{\theta}Y_{\ell m}-
    \omega H_{\ell}(\omega r)v_{\ell m}\frac{mY_{\ell m}}{\sin\theta}
    \right)\,,
\\ \nonumber\\
\label{sol_wz_mf3}
&& B^{\hat\phi}  =  {\rm i}\frac{\,{\rm e}^{-{\rm i}\omega t}}
    {\sqrt{\ell(\ell+1)}}
    \left(DH_{\ell}(\omega r) u_{\ell m}\frac{mY_{\ell m}}{\sin\theta}
    -\omega H_{\ell}(\omega r)v_{\ell m}\,\partial_{\theta}Y_{\ell m}\right)\,,
\end{eqnarray}
while the electric field components are expressed as [see equations
  100--102 of paper I]
\begin{eqnarray}
\label{sol_wz_ef1}
&& E^{\hat r}  =  -\frac{\,{\rm e}^{-{\rm i}\omega t}
    \sqrt{\ell\left(\ell+1\right)}}{r}
    H_{\ell}(\omega r)v_{\ell m}Y_{\ell m}\,,
\\\nonumber\\
\label{sol_wz_ef2}
&& E^{\hat\theta}  =  -\frac{\,{\rm e}^{-{\rm i}\omega t}}
    {\sqrt{\ell(\ell+1)}}
    \left(DH_{\ell}(\omega r)v_{\ell m}\,\partial_{\theta}Y_{\ell m}+
    \omega H_{\ell}(\omega r)u_{\ell m}\frac{mY_{\ell m}}{\sin\theta}
    \right)\,,
\\ \nonumber\\
\label{sol_wz_ef3}
&& E^{\hat\phi}  =  -{\rm i}\frac{\,{\rm e}^{-{\rm i}\omega t}}
    {\sqrt{\ell(\ell+1)}}
    \left(DH_{\ell}(\omega r) v_{\ell m}\frac{mY_{\ell m}}{\sin\theta}
    +\omega H_{\ell}(\omega r)u_{\ell m}\,\partial_{\theta}Y_{\ell m}
    \right)\,.
\end{eqnarray}

Expressions (\ref{sol_wz_mf1})--(\ref{sol_wz_ef3}) have the same
functional form as in the Newtonian limit, but general-relativistic
corrections are introduced through the integration coefficients $u_{\ell
  m}$ and $v_{\ell m}$, which are specified through the matching of the
electromagnetic fields (\ref{sol_wz_mf1})--(\ref{sol_wz_ef3}) at the
stellar surface as [see equations 105 and 106 of paper I]
\begin{eqnarray}
\label{v_coef}
&& v_{\ell m} = \frac{1}{\sqrt{\ell(\ell+1)}}
    \frac{\,{\rm e}^{{\rm i}\omega_{_\textrm{R}}t}}{DH_{\ell}(\omega_{_\textrm{R}}R)N_{_\textrm{R}}}
    \int {\rm d}\Omega\left\{\partial_{\theta}Y^*_{\ell m}
    \left[\delta v_{_\textrm{R}}^{\hat r} B^{\hat \phi}_{_\textrm{R}}
        -\delta v_{_\textrm{R}}^{\hat \phi}B^{\hat r}_{_\textrm{R}}\right] +
        {\rm i}\frac{mY^*_{\ell m}}{\sin\theta}
        \left[\delta v_{_\textrm{R}}^{\hat\theta} B^{\hat r}_{_\textrm{R}}
        -\delta v_{_\textrm{R}}^{\hat r}B^{\hat \theta}_{_\textrm{R}}\right]\right\}\,,
\\ \nonumber\\
\label{u_coef}
&& u_{\ell m} = \frac{1}{\sqrt{\ell(\ell+1)}}
    \frac{\,{\rm e}^{{\rm i}\omega_{_\textrm{R}}t}}{H_{\ell}(\omega_{_\textrm{R}}R)N_{_\textrm{R}}\omega_{_\textrm{R}}}
    \int {\rm d}\Omega\left\{{\rm i}\,\partial_{\theta}Y^*_{\ell m}
    \left[\delta v_{_\textrm{R}}^{\hat r} B^{\hat \theta}_{_\textrm{R}}
        -\delta v_{_\textrm{R}}^{\hat \theta}B^{\hat r}_{_\textrm{R}}\right] +
        \frac{mY^*_{\ell m}}{\sin\theta}
        \left[\delta v_{_\textrm{R}}^{\hat\phi} B^{\hat r}_{_\textrm{R}}
        -\delta v_{_\textrm{R}}^{\hat r}B^{\hat\phi}_{_\textrm{R}}\right]\right\}\,.
\end{eqnarray}
In what follows we will discuss the expressions for the electromagnetic
fields in the wave-zone which are produced by the different velocity
fields discussed in Sections \ref{spheroid} and \ref{to}.

\subsection{Electromagnetic fields produced by spheroidal oscillations}
\label{spheroid_wz}

The oscillating electric and magnetic fields will obviously produce
electromagnetic waves and the outgoing electromagnetic radiation in the
case of spheroidal oscillations will need to satisfy the conditions that
$m=m'$ and $\ell=\ell', \ell'\pm 1$  \citep[see, \eg][]{Rose1955}. For
typical stellar spheroidal oscillations \citep{Duncan1998}, it is easy to
estimate that $\omega_{_\textrm{R}}R\ll 1$ at least for the low-order modes, and
in this limit we can calculate, for example, the electromagnetic fields
radiated by an axisymmetric dipolar oscillation (\ie with $\ell'=1,
m'=0$). In this specific case, the non-vanishing coefficients are given by
$u_{10}, u_{20}, v_{20}$ and have explicit expressions as
\begin{eqnarray}
\label{u_20_sph_wz} && u_{10}=-{\rm i}\frac{3}{8\sqrt{2}}
    \frac{\omega_{_\textrm{R}} R^2}{N_{_\textrm{R}}}B_{_0}
    \left(\eta_{_\textrm{R}}h_{_\textrm{R}}-3\xi_{_\textrm{R}}f_{_\textrm{R}}\right) \sin\chi\,,
\ \nonumber\\ \nonumber\\
&& u_{20}=\frac{1}{3}\sqrt{\frac{2}{5}}
    \frac{\omega^2_{_\textrm{R}} R^3}{N_{_\textrm{R}}}B_{_0}
    \left(\eta_{_\textrm{R}}h_{_\textrm{R}}+\xi_{_\textrm{R}}f_{_\textrm{R}}\right) \cos\chi\,,
\qquad
v_{20}=-\frac{1}{16}\sqrt{\frac{5}{2}}
    \frac{\omega^3_{_\textrm{R}} R^4}{N_{_\textrm{R}}}B_{_0}
    \eta_{_\textrm{R}}h_{_\textrm{R}} \sin\chi\,,
\end{eqnarray}
so that the electromagnetic fields (\ref{sol_wz_mf1})--(\ref{sol_wz_ef3})
induced in the wave zone are given as the real parts of the following
solutions
\begin{eqnarray}
\label{wz_spheroid_1_1}
&& B^{\hat r} =
    \frac{1}{\sqrt{12\rm{\pi}}}\frac{R^2}{N_{_\textrm{R}}r^2}
    B_{_0}\left[\frac{\omega_{_\textrm{R}}R}{N_{_\textrm{R}}}
    \left(\eta_{_\textrm{R}}h_{_\textrm{R}}+\xi_{_\textrm{R}}f_{_\textrm{R}}\right)
    (3\cos^2\theta -1)\cos\chi +
    {\rm i}\frac{9}{8}
    \left(\eta_{_\textrm{R}}h_{_\textrm{R}}-3\xi_{_\textrm{R}}f_{_\textrm{R}}\right)
        \sin\chi\cos\theta\right] {\,{\rm e}^{{\rm i}\omega (r-t)}}\,,
\\ \nonumber\\
\label{wz_spheroid_1_2}
&& B^{\hat\theta} =
    \frac{1}{\sqrt{12\rm{\pi}}}\frac{\omega_{_\textrm{R}}R^2}{N_{_\textrm{R}}r}
    B_{_0}\left[\omega_{_\textrm{R}}R
    \left(\eta_{_\textrm{R}}h_{_\textrm{R}}+\xi_{_\textrm{R}}f_{_\textrm{R}}\right)
    \sin\theta\cos\theta
        \cos\chi  +
    \frac{9}{16}\left(\eta_{_\textrm{R}}h_{_\textrm{R}}-3\xi_{_\textrm{R}}f_{_\textrm{R}}\right)
        \sin\chi\sin\theta\right] {\,{\rm e}^{{\rm i}\omega (r-t)}}\,,
\\ \nonumber\\
\label{wz_spheroid_1_3}
&& B^{\hat\phi} = \frac{15}{32}\frac{1}{\sqrt{12\rm{\pi}}}
    \frac{\omega_{_\textrm{R}}^3R^4}{N_{_\textrm{R}}r}
    B_{_0} \eta_{_\textrm{R}}h_{_\textrm{R}}
        \sin\chi\sin\theta\cos\theta {\,{\rm e}^{{\rm i}\omega (r-t)}} \,,
\\ \nonumber\\
\label{wz_spheroid_1_4} && E^{\hat r} = {\rm
i}\frac{15}{32}\frac{1}{\sqrt{12\rm{\pi}}}
    \frac{\omega_{_\textrm{R}}^3R^4}{N_{_\textrm{R}}r}
    B_{_0} \eta_{_\textrm{R}}h_{_\textrm{R}}
        \sin\chi(3\cos^2\theta-1) {\,{\rm e}^{{\rm i}\omega (r-t)}}\,,
\\ \nonumber\\
\label{wz_spheroid_1_5}
&& E^{\hat\theta} = \frac{15}{32}\frac{1}{\sqrt{12\rm{\pi}}}
    \frac{\omega_{_\textrm{R}}^3R^4}{N_{_\textrm{R}}r}
    B_{_0} \eta_{_\textrm{R}}h_{_\textrm{R}}
        \sin\chi\sin\theta\cos\theta {\,{\rm e}^{{\rm i}\omega (r-t)}} \,,
\\ \nonumber\\
\label{wz_spheroid_1_6}
&& E^{\hat\phi} =
    -\frac{1}{\sqrt{12\rm{\pi}}}\frac{\omega_{_\textrm{R}}R^2}{N_{_\textrm{R}}r}
    B_{_0}\left[\omega_{_\textrm{R}}R
    \left(\eta_{_\textrm{R}}h_{_\textrm{R}}+\xi_{_\textrm{R}}f_{_\textrm{R}}\right)
    \sin\theta\cos\theta
        \cos\chi  +
    \frac{9}{16}\left(\eta_{_\textrm{R}}h_{_\textrm{R}}-3\xi_{_\textrm{R}}f_{_\textrm{R}}\right)
        \sin\chi\sin\theta\right] {\,{\rm e}^{{\rm i}\omega (r-t)}}\,.
\end{eqnarray}
Note that, in the expressions above, we have omitted the symbol $\delta$
for the magnetic and electric fields in the wave-zone because the fields
produced there are exclusively the ones due to the stellar
perturbation. Since the wave-zone is located well outside the light
cylinder, \ie at $r \gg r_{\textrm{lc}} := 1/\Omega$, expressions
(\ref{wz_spheroid_1_1})--(\ref{wz_spheroid_1_6}) show that, in this region,
the electromagnetic fields behave essentially as radially outgoing waves,
for which $|B^{\hat r}/B^{\hat \theta}| \sim |B^{\hat r}/B^{\hat \phi}|
\sim 1/\omega r \ll 1$. We postpone to Appendix~\ref{em_wz_spheroidal}
the presentation of the electromagnetic fields produced by other
higher order spheroidal modes.

\subsubsection{Electromagnetic fields produced by radial
oscillations} \label{ro_wz}

From the asymptotic forms (\ref{small}) and (\ref{large}), it follows
that the solutions (\ref{sol_wz_mf1})--(\ref{sol_wz_ef3}) in the
wave-zone for radial oscillations (\ref{rad_puls}) and for a dipolar
perturbation are given by
\begin{eqnarray}
\label{rad_wz_mf1}
&& B^{\hat r} = \frac{2h_{_\textrm{R}}R^2}{N_{_\textrm{R}}^2r^2}B_{_0}
    \eta_{_\textrm{R}}\,{\rm e}^{{\rm i}\omega (r-t)}
    (\cos\chi\cos\theta +\sin\chi\sin\theta \,{\rm e}^{{\rm i}\phi}) \,,
\\ \nonumber\\
\label{rad_wz_mf2} && B^{\hat\theta} = -{\rm i}
\frac{h_{_\textrm{R}}\omega_{_\textrm{R}}R^2}{N_{_\textrm{R}}r}
    B_{_0}\eta_{_\textrm{R}}\,{\rm e}^{{\rm i}\omega (r-t)}
    (\cos\chi\sin\theta - \sin\chi\cos\theta \,{\rm e}^{{\rm i}\phi}) \,,
\\ \nonumber\\
\label{rad_wz_mf3}
&& B^{\hat\phi} = - \frac{h_{_\textrm{R}}\omega_{_\textrm{R}}R^2}{N_{_\textrm{R}}r}
    B_{_0}\eta_{_\textrm{R}}\,{\rm e}^{{\rm i}\omega (r-t)}\sin\chi \,{\rm e}^{{\rm i} \phi}  \,,
\\ \nonumber\\
\label{rad_wz_ef2}
&& E^{\hat\theta} = - \frac{h_{_\textrm{R}}\omega_{_\textrm{R}}R^2}{N_{_\textrm{R}}r}B_{_0}
        \eta_{_\textrm{R}}\,{\rm e}^{{\rm i}\omega (r-t)}\sin\chi \,{\rm e}^{{\rm i}\phi} \,,
\\ \nonumber\\
\label{rad_wz_ef3} && E^{\hat\phi} = {\rm i}
\frac{h_{_\textrm{R}}\omega_{_\textrm{R}}R^2}{N_{_\textrm{R}}r}
    B_{_0}\eta_{_\textrm{R}}\,{\rm e}^{{\rm i}\omega (r-t)}
    (\cos\chi\sin\theta-\sin\chi\cos\theta\cos\phi) \,,
\end{eqnarray}
where the non-vanishing integration constants have explicit expressions
\begin{equation}
\label{u_10_rad_wz}
u_{10}=-2\sqrt{\frac{2\rm{\pi}}{3}}
    \frac{h_{_\textrm{R}} R^2}{N_{_\textrm{R}}}B_{_0}
    \eta_{_\textrm{R}} \cos\chi\,, \qquad \qquad
u_{11}=2\sqrt{\frac{4\rm{\pi}}{3}}
    \frac{\omega_{_\textrm{R}}h_{_\textrm{R}} R^2}{N_{_\textrm{R}}}B_{_0}
    \eta_{_\textrm{R}} \sin\chi\,.
\end{equation}
These electromagnetic field components satisfy the force-free condition
$E^{\hat i}B_{\hat i}=0$ and in the Newtonian limit they reduce to the
results of \citet{Muslimov1986} [\cf equations B3 and B4 of
  \citet{Muslimov1986}].

\subsection{Electromagnetic fields produced by toroidal oscillations}
\label{to_wz}

In full analogy with the case of spheroidal oscillations, the outgoing
electromagnetic radiation in the case of toroidal oscillations needs to
satisfy the conditions that $m=m'$ and $\ell=\ell', \ell'\pm 1$
  \citep[see, \eg][]{Rose1955} For typical stellar toroidal oscillations, which have
only azimuthal velocities \citep[see, \eg][]{Duncan1998}, $\omega_{_\textrm{R}}R\ll
1$ for the low-order modes, and in this limit we can calculate explicitly
the electromagnetic fields radiated by the mode with $\ell'=1, m'=0$ as
\begin{eqnarray}
&& B^{\hat\phi} =
    \frac{\rm i}{4\sqrt{3\rm{\pi}}}
        \frac{f_{_\textrm{R}}\omega^3_{_\textrm{R}}R^4}{N_{_\textrm{R}}r}B_{_0}
    \eta_{_\textrm{R}}{\,{\rm e}^{{\rm i}\omega (r-t)}}
    \sin\theta\cos\theta\cos\chi \,,
\\ \nonumber\\
&& E^{\hat r} =
    -\frac{1}{4\sqrt{3\rm{\pi}}}
        \frac{f_{_\textrm{R}}\omega^2_{_\textrm{R}}R^4}{ N_{_\textrm{R}}^2r^2}B_{_0}
    \eta_{_\textrm{R}}{\,{\rm e}^{{\rm i}\omega (r-t)}}
    \left(3\cos^2\theta -1\right)\cos\chi \,,
\\ \nonumber\\
&& E^{\hat\theta} =
    \frac{{\rm i}}{4\sqrt{3\rm{\pi}}}
        \frac{f_{_\textrm{R}}\omega^3_{_\textrm{R}}R^4}{N_{_\textrm{R}}r}B_{_0}
    \eta_{_\textrm{R}}{\,{\rm e}^{{\rm i}\omega (r-t)}}
    \sin\theta\cos\theta\cos\chi \,,
\end{eqnarray}
where the only non-vanishing integration constant is given by
\begin{equation}
\label{v_20_tor_wz} v_{20}=-\frac{{\rm i}}{15}\sqrt{\frac{5}{2}}
    \frac{f_{_\textrm{R}}\omega^3_{_\textrm{R}} R^4}{N_{_\textrm{R}}}B_{_0}
    \eta_{_\textrm{R}} \cos\chi\,.
\end{equation}
Also in this case, the Newtonian limits of the expressions above reduce
to the expressions that can be derived from the general equations 25 and
26 of \citet{Muslimov1986}. We postpone to Appendix~\ref{em_wz_toroidal}
the presentation of the electromagnetic fields produced by other higher
order toroidal modes.

\section{Electromagnetic damping of oscillations}
\label{oscilltn_damping}

As mentioned in the Introduction, the scope of this Section is to compute
the damping times associated with the most typical modes of oscillation
when the latter produce pure electromagnetic waves carrying away the
energy stored in the oscillations. Of course, if an oscillation is
excited in a compact star, there will be also other mechanisms, most
notably gravitational-wave emission, that, together with the
electromagnetic ones, will combine to damp the stellar
oscillations\footnote{Viscous damping is also expected to be present but
  would act mostly on very high-order modes and on longer time-scales. We
  recall, in fact, that in Newtonian physics, the viscous oscillation
  damping time-scale can be estimated as \citep{Chugunov2005} $\tau_{_{\rm
      visc}} \sim \varepsilon/\dot{\varepsilon}$, where $\varepsilon \sim
  \rho v^2$ is the energy density of the oscillation averaged over a
  period, with $\rho$ and $v$ being respectively the characteristic
  density and velocity, and $\dot{\varepsilon} \sim
  \eta\left({v}/{\lambda}\right)^2$ the viscous-loss rate, with $\lambda
  \sim R/\ell$ and $\eta$ the shear viscosity. As a result, $\tau_{_{\rm
      visc}} \sim \lambda^2 {\rho}/{\eta}$, so that the characteristic
  damping time-scales decreases for higher modes.}. For simplicity, we
will consider these mechanisms as acting independently and compute in the
following Sections the electromagnetic radiative losses for the classes
of oscillations considered so far and compare them with the losses via
gravitational radiation.

The energy lost via electromagnetic radiation, $L_{_{\rm EM}}$, can be
readily calculated through the integral of the radial component of the
Poynting vector $\vec{\boldsymbol{P}}$
\begin{equation}
\label{power} 
L_{_{\rm{EM}}} := \int_{\partial \Sigma} P^{\hat r}{\rm d}S =
    \frac{1}{4\rm{\pi}} \int_{\partial \Sigma}
    \left(\vec{{\boldsymbol{E}}}\times
    \vec{\boldsymbol{B}} \right)^{\hat r} {\rm d}S\,,
\end{equation}
which is integrated over the two-sphere $\partial\Sigma$ of radius $r\gg
1/\Omega > R$ and the surface element ${\rm d}S$ (the integration is
taken on the surface $\partial\Sigma$ at distance $r\gg R$ where the
space-time can be reasonably approximated as flat). Substituting in
(\ref{power}) the general expressions for the electromagnetic radiation
(\ref{sol_wz_mf2})--(\ref{sol_wz_mf3}) and
(\ref{sol_wz_ef2})--(\ref{sol_wz_ef3}), we can easily find that the
oscillation energy loss due to the electromagnetic radiation is simply
expressed in terms of the integration constants $u_{\ell m}$ and $v_{\ell
  m}$ [see equation 148 of paper I and its discussion for more details],
\ie
\begin{equation}
\label{em_energy_loss}
L_{_{\rm EM}} = \frac{1}{8\rm{\pi}}
    \left(\left|u_{\ell m}\right|^2 +
    \left|v_{\ell m}\right|^2\right)
    \,,
\end{equation}
and thus depends on the value of the gravitational compactness parameter
$M/R$ appearing in the integration constants $u_{\ell m}$ and $v_{\ell
  m}$.

\subsection{Damping due to electromagnetic radiation from
toroidal oscillations}

Toroidal modes, which preserve the stellar shape at the lowest order, are
particularly interesting to estimate the electromagnetic losses since
these modes are expected to follow from large crustal fractures and
starquakes, and may be especially easy to excite on longer time-scales
because the restoring force is due to the relatively weak Coulomb forces
of the crustal ions. Furthermore, toroidal shear deformations require
much less energy than do radial deformations of the same amplitude and
the damping rate is expected to be proportional to the oscillation
frequency, which is small for low-order toroidal modes.

Before considering the general-relativistic expression for the
electromagnetic losses of the most interesting modes, we recall that the
corresponding Newtonian expression for the power $L_{_{\rm
    EM}}(\ell',m')\vert_{\rm Newt}$ radiated by a toroidal oscillation
mode with $(\ell',m')$ with $\ell'>1$ is given by \citep{McDermott1984}
\begin{equation}
\label{em_radiation_toroidal_newt}
L_{_{\rm EM}}(\ell',m')\vert_{\rm Newt}=\frac{c}{8\rm{\pi}}
        \eta^2_{_\textrm{R}}\left(\frac{\omega_{_\textrm{R}}R}{c}\right)^{2\ell'}
        \frac{B_{_0}^2 R^2}{\ell'\left(2\ell'-1\right)
        \left[\left(2\ell'-3\right)!!\right]^2} \left[\frac{m^{\prime
        2}}{\left(2\ell'-1\right)\left(\ell'+1\right)}
        +\frac{\left(\ell'+1\right)^2\left(\ell^{\prime 2}-
        {m}^{\prime 2}\right)}
        {\left(\ell'-1\right)\left(2\ell'+1\right)}\right] \cos^2\chi
        \,,
\end{equation}
where we have assumed that $\omega_{_\textrm{R}} R \ll 1$, which is surely valid
at least for the lowest-order modes. Expression
(\ref{em_radiation_toroidal_newt}) can be extended to general relativity
either by a proper calculation of the integration constants or, more
simply, by considering that the general-relativistic corrections will
come in the form of an amplification factor of order ${f_{_\textrm{R}}}/{N_{_\textrm{R}}}$
for the magnetic-field strength and of a frequency increase of the order
$\left(1/N_{_\textrm{R}}\right)^{2\ell'}$ to compensate for the gravitational
redshift. Of course the two routes lead to the same answer and thus to a
radiated power
\begin{eqnarray}
\label{em_radiation_toroidal}
&&L_{_{\rm EM}}(\ell',m') =
\left(\frac{f_{_\textrm{R}}}{N_{_\textrm{R}}} \right)^2
\left(\frac{1}{N_{_\textrm{R}}}\right)^{2 \ell'}L_{_{\rm EM}}(\ell',m')\vert_{\rm Newt}
\nonumber \\
&&\hskip 1.7cm = \frac{c}{8\rm{\pi}}
    \left[
    \frac{f_{_\textrm{R}}}{N_{_\textrm{R}}}
        \left(\frac{\omega_{_\textrm{R}}R}{c}\right)^{\ell'}
        \right]^2
    \frac{\eta^2_{_\textrm{R}} B_{_0}^2 R^2}{\ell'\left(2\ell'-1\right)
    \left[\left(2\ell'-3\right)!!\right]^2}
    \left[\frac{m^{\prime 2}}{\left(2\ell'-1\right)\left(\ell'+1\right)}
    +\frac{\left(\ell'+1\right)^2\left(\ell^{\prime 2}-
    {m}^{\prime 2}\right)}
    {\left(\ell'-1\right)\left(2\ell'+1\right)}\right]
    \cos^2\chi \,.
\end{eqnarray}

\begin{figure}
\begin{center}
\label{figure1}
\includegraphics[width=8.0cm,angle=0]{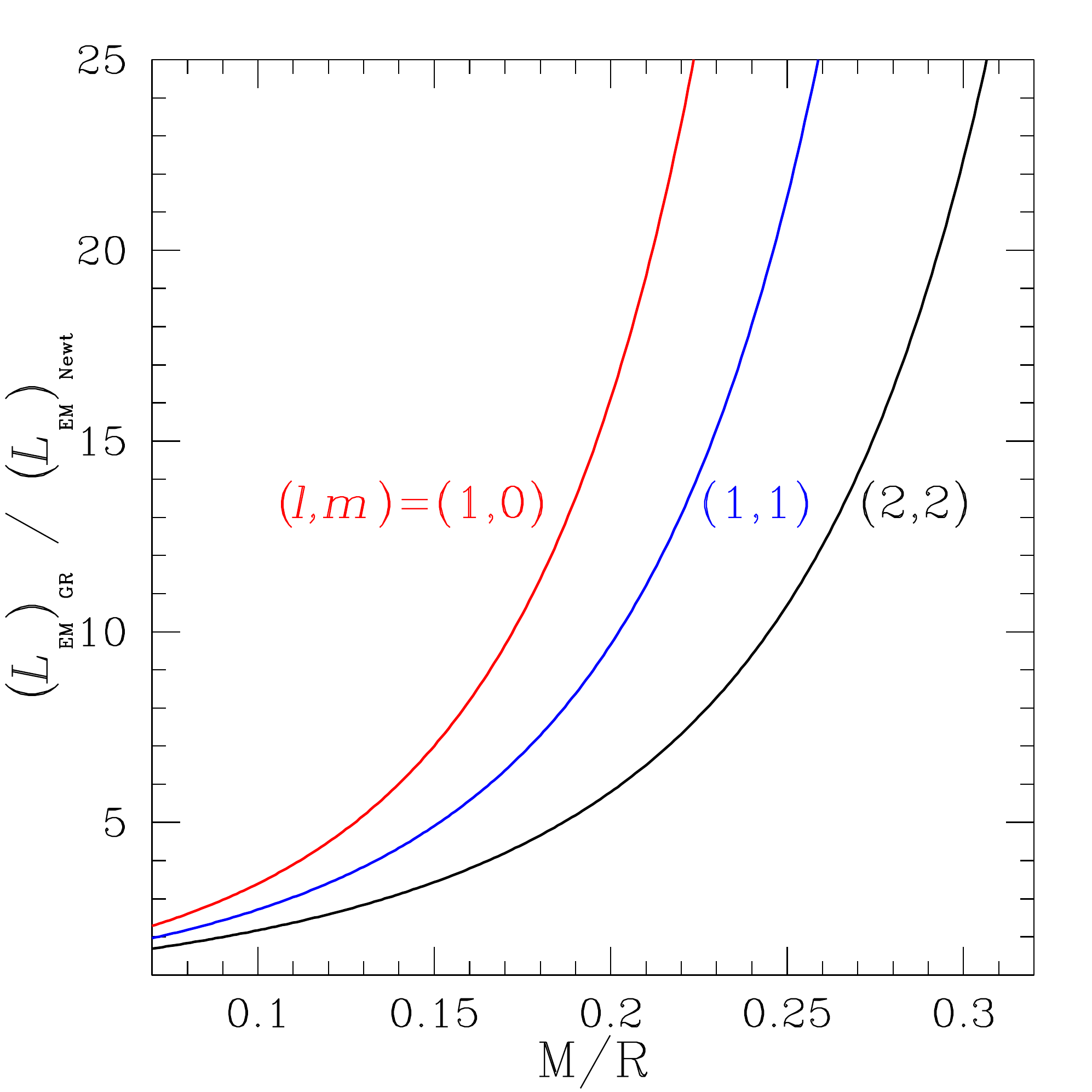}
\caption{General-relativistic amplification of the energy loss through
  multipolar electromagnetic radiation coming from toroidal oscillations
  and shown as a function of the stellar compactness [\cf
    Eq. \eqref{em_radiation_toroidal}]. Different lines correspond to
  different multipolar orders $\ell'$; note that the general-relativistic
  corrections increase the losses of one order of magnitude or
  more.} \label{fig1}
\end{center}
\end{figure}

Fig.~\ref{figure1} shows the dependence on the stellar compactness of
the energy emission given by expression (\ref{em_radiation_toroidal}) for
a few representative modes and it is apparent that the corrections can be
rather large and at least of one order of magnitude for typical neutron
stars with a surface magnetic field $B_{\rm s}=10^{12} \,{\rm
  G}$\footnote{Magnetic fields of this strength do not produce
  significant changes in the mode frequencies \citep{Lee2007}, so that we
  can safely ignore such corrections.}. As a reference, the
electromagnetic luminosity produced by a purely quadrupolar mode (\ie
with $\ell'=2, m'=2$) can be expressed as
\begin{equation}
\label{L_22_toroid}
L_{_{\rm EM}}(2,2) \approx
   6\times 10^{36}
   \left(\frac{\eta_{_\textrm{R}}}{\omega R}\right)^2
   \frac{f_{_\textrm{R}}^2}{N^8_{_\textrm{R}}}
   \, B_{12}^2
   \, R_{6}^{8}
   \, \omega_3^{6} \cos^2\chi
   \quad {\rm erg/s} \,,
\end{equation}
where $B_{12} := {B_{\rm s}}/({10^{12}\,{\rm G}})$, $\omega_{3} := \omega
/(10^3\,{\rm rad/s})$, $R_{6} := {R}/({10^{6}\,{\rm cm}})$.

Given $E_{_\textrm{T}}$ as the kinetic energy contained in the stellar
oscillations and assuming that all of it is lost to the emission of
electromagnetic waves, we can define the electromagnetic decay time-scale
of the $(\ell, m)$ mode to be
\begin{equation}
\label{tau_em}
\tau_{_{\rm EM}}^{^{\rm GR}}(\ell,m) :=
\frac{2E_{_\textrm{T}}}{L_{_{\rm EM}}^{^{\rm GR}}(\ell,m)}\,,
\end{equation}
where the factor of 2 on the right-hand side of expression (\ref{tau_em}) is
introduced because of the averaging over one oscillation period. For
simplicity, we compare the electromagnetic time-scales $\tau_{_{\rm
    EM}}^{^{\rm GR}}$ with the gravitational-radiation ones as computed
by used by \citet{McDermott1988} in their model calculations denoted as
NS13T8. Such a model has a mass $M = 1.326 \,M_{\odot}$, a radius
$R=7.853 \, {\rm km}$, and magnetic field $B=10^{12} \,{\rm G}$. While
this model can no longer be considered realistic (the radius is far
smaller than what expected from more modern equations of state), it has
been used also quite recently \citep{Messios2001,YoshidaLee2002,Lee2007}
and it is useful here as for this model \citet{McDermott1988} have
provided a complete list of eigenfrequencies, luminosities and damping
times for a number of modes\footnote{More recent but also more restricted
  information can be found in the works of
  \citet{Gaertig:2008a,Gaertig2011}, who have studied the oscillations of
  rapidly rotating stars.}. As for the other models in
\citet{McDermott1988}, NS13T8 was obtained from the fully
general-relativistic calculations of neutron stars by
\citet{Richardson1982}. The outer crust extended down to the neutron-drip
point at $\rho = 4.3 \times 10^{11} {\rm g \, cm^{-3}}$ and were assumed
to consist of bare iron nuclei embedded in a uniform, neutralizing,
degenerate electron gas. The inner crust extended from the neutron-drip
point to the base of the crust at $\rho = 2.4 \times 10^{14} {\rm g \,
  cm^{-3}}$, and it was assumed to consist of nuclei with atomic number
$Z \sim 40$, degenerate electrons, and degenerate, nonrelativistic
neutrons. At the greater densities, the lattice was assumed to dissolve
and the core of the neutron star was taken to consist of a mixture of
free and highly degenerate neutrons, protons and electrons. Finally,
model NS13T8 was assumed to have a solid crust with relative thickness
$\Delta r/R \sim 0.055$ and a surface ocean with relative thickness of
the $\Delta r/R \sim 2.3 \times 10^{-3}$.

Table~\ref{toroid_damping_time} reports the different damping times for
the first toroidal modes for model NS13T8 with a total time-averaged
kinetic energy in the crust defined as
\begin{equation}
E_{_{\rm T}} = \frac{1}{2}\omega^2 \ell(\ell+1)
\int_{{\rm crust}}\rho\,\tilde{\eta}^2r^2dr\,.
\end{equation}
The different columns in the table report the main characteristics of the
oscillation modes, such as the frequency and kinetic energy, the power of
electromagnetic radiation, the gravitational and electromagnetic damping
times in the Newtonian and general-relativistic cases. The Newtonian
expressions for the pulsation frequencies, the kinetic energy and the
damping times are those reported in Tables 4 and 6 of
\citet{McDermott1988} and reproduced in Table~\ref{toroid_damping_time}
in columns 1--3 and 5--6. The general-relativistic values of the
electromagnetic luminosity and of the damping time reported in columns 4
and 7 are those calculated via expression
(\ref{em_radiation_toroidal}). The last two columns represent the ratio
of the gravitational and electromagnetic time-scales and the ratio of the
electromagnetic time-scales in the Newtonian and general-relativistic
case, respectively. Of course, the damping times $\tau^{^{\rm GR}}_{_{\rm
    EM}}$ are much more interesting when compared with the corresponding
damping times $\tau_{_{\rm GW}}$ calculated when the kinetic energy
$E_{_{\rm T}}$ is instead lost to gravitational waves. These were
computed by \citet{Schumaker1983} for the first-order perturbations in
the displacement functions of a fully general-relativistic stellar model
and are obviously reported for modes with $\ell \geq 2$, since no
gravitational radiation can be produced by dipolar oscillations.

Early calculations of toroidal oscillations assumed the 'free-slip'
condition of the solid crust over the fluid core \citep{Hansen1980,
  McDermott1988}, but these models have been recently improved to include
the gravitational redshift, the increase in the shear modulus due to
magnetic pressure, more realistic models of crust composition and
elasticity. In particular, more recent calculations of \citet{Duncan1998}
and \citet{Piro2005} estimate the redshifted frequency $\nu :=
\omega/2\rm{\pi}$ of the fundamental $\ell=2$ toroidal mode\footnote{A
  toroidal mode with $\ell=1$ is not allowed as it would violate
  angular-momentum conservation.} through the empirical expression
\begin{equation}
\label{nu2t0}
\nu(_2t_0)=29.8\frac{(1.71 - 0.71\,M_{1.4}\,R_{6}^{-1})^{1/2}}
   {R_{6}(0.87+0.13\,M_{1.4}\,R_{6}^{-2})}
   \left[1+\left(\frac{B}{B_{\rm s}}\right)^{2}\right]^{1/2}
   \qquad {\rm Hz} \,,
\end{equation}
while the frequencies of modes of order higher than $n=0$ are simply
given by
\begin{equation}
\nu(_\ell t_0) = \nu(_2t_0)
\left[\frac{\ell(\ell+1)}{6}\right]^{1/2}
\,,
\end{equation}
with $M_{1.4} := M/1.4 \,M_\odot$\footnote{\citet{Duncan1998} also
  estimated the kinetic energy of the $_2t_0$-mode as
  $E_{_{T}}(_2t_0)=5\times 10^{47} ({\eta_{_\textrm{R}}}/{R})^{2}
  M_{1.4}^{-1}R_{6}^{4} \left(0.77+0.23M_{1.4}R_{6}^{-2}\right)$ erg.}.

Using $B_{\rm s} \approx 4\times 10^{15} \,{\rm G}$, expression
(\ref{nu2t0}) gives, \eg $\nu(_2t_0)=28.5 \,{\rm Hz}$, which is the
general-relativistic value of the Newtonian value of $57.7 \, {\rm Hz}$
previously calculated by \citet{McDermott1988} for the slightly different
NS13T8 model. The difference between the Newtonian and the
general-relativistic values is partly due to the redshift factor
$(1-2M/R)^{1/2}$, which is approximately $0.8$ at the surface of the star
[\cf equation \eqref{redshift}]. It is necessary to underline that it is
reasonable to assume the oscillation modes to be independent of the
magnetic field strength for $B\leq 10^{15} \,{\rm G}$, a limit which is
several orders of magnitude greater than the dipolar magnetic fields at
the surface of the typical neutron stars \citep{Piro2005}. Axisymmetric
toroidal modes of magnetized neutron stars calculated in the
general-relativistic Cowling approximation can be also found in the work
of \citet{Asai2014}.

\begin{table}
\label{toroid_damping_time}
\caption{Damping times of toroidal modes for NS13T8 model of a neutron
  star \citep{McDermott1988}. The different columns report
  characteristics of oscillation modes as frequency and kinetic energy,
  power of electromagnetic radiation, gravitational and electromagnetic
  damping times in the Newtonian and general-relativistic cases. The last
  column represents the ratio of the gravitational and electromagnetic
  time-scales. For the various nodes we use the standard notation in
  stellar seismology so that, for instance, the mode $_{l}t_n$ refers to
  a toroidal $t$ mode with harmonic index $\ell=l$ and overtone number
  $n$; for simplicity we limit ourselves to the first overtone. Finally
  note that although large, the kinetic energies in the toroidal
  oscillations estimated by \citet{McDermott1988} are still only a small
  fraction of the binding energy available in the system, \ie $\gtrsim
  10^{54}\,{\rm erg}$. }
\begin{tabular}{|c|c|l|l|l|l|l|l|l|c}
\hline
${\rm Mode}$ & $\omega$  & $E_{_{\rm T}}$
&  $L^{^{\rm Newt}}_{_{\rm EM}} $&
$L^{^{\rm GR}}_{_{\rm EM}} $&
$\tau_{_{\rm GW}}$ & $\tau^{^{\rm Newt}}_{_{\rm EM}}$ &
$\tau^{^{\rm GR}}_{_{\rm EM}}$ &
$\tau_{_{\rm GW}}$/$\tau^{^{\rm GR}}_{_{\rm EM}}$ & $\tau^{^{\rm Newt}}_{_{\rm EM}}/\tau^{^{\rm GR}}_{_{\rm EM}}$\\
& (kHz) & (erg) & (erg/s) & (erg/s) & (s) & (s) & (s) & & \\
\hline
$_1t_{1}$ & $17.9$ & $1.09\times 10^{49}$ & $1.77\times
10^{43}$ & $1.57\times 10^{44}$ & $-$ & $1.23\times 10^{6}$ &
$1.39\times 10^{5}$ & $-$& $8.85$
\\
$_2t_{0}$ & $\phantom{0}0.4$ & $3.31\times 10^{47}$ & $6.86\times 10^{32}$ & $3.45\times 10^{33}$ & $6.62\times
 10^{11}$
& $9.65\times 10^{14}$ & $1.92\times 10^{14}$ & $< 10^{-3}$ &
$5.03$
\\
 $_2t_{1}$ & $17.9$ & $3.26\times 10^{49}$ & $9.32\times 10^{42}$ & $4.96\times 10^{43}$ & $7.60\times 10^{5}$ &
$7.00\times 10^{6}$ & $1.31\times 10^{6}$ & $\phantom{0}0.6$ & $5.34$
\\
  \hline
\end{tabular}
\end{table}

Overall, the values reported in Table~\ref{toroid_damping_time} reveal
that low-frequency quadrupolar modes are damped more efficiently by
gravitational radiation than by electromagnetic radiation. In contrast,
high-frequency quadrupolar modes are damped more efficiently by
electromagnetic radiation. As two representative examples, the ratio
$\tau_{_{\rm GW}}/\tau^{^{\rm GR}}_{_{\rm EM}}$ is approximately
$3.5\times 10^{-3}$ for the lowest quadrupolar toroidal model $_2t_0$,
while it grows to approximately $10^2$ for the fourth overtone
$_2t_4$. This scaling is very interesting as it indicates that an
oscillating neutron star subject to toroidal oscillations will rapidly
lose most of its kinetic energy to the lowest order quadrupolar
oscillations, but will continue to emit electromagnetic energy for a
longer time-scale in terms of its higher-order quadrupolar oscillations.

It is also important to underline that the damping times reported in
Table~\ref{toroid_damping_time} refer to a fiducial neutron star with a
magnetic field $B_{\rm s}=10^{12}\,{\rm G}$ and that the electromagnetic
damping time scales like $B_{\rm s}^{-2}$ [\cf
  equations~(\ref{em_radiation_toroidal_newt}) and~(\ref{tau_em})]. As a
result, the damping times reported would need to be modified
significantly, \ie of about four to six orders of magnitude, if the oscillations
are taking place in a magnetar with the surface magnetic field $B_{\rm
  s}=10^{14}-10^{15}\,{\rm G}$.

As a concluding remark, and to stress that the harmonic electromagnetic
variability discussed above is within the range of present observations,
we note that \citet{Clemens2004, {Clemens2008}, Rosen2008} have proposed
a detailed analysis of the evolution of the pulse shapes of radio pulsars
due to high-order oscillations with multipole numbers as large as $\ell
\sim 70$.  This interpretation is difficult to conciliate with our
results, which indicate that the general-relativistic redshift
corrections would be very large for such high multipoles, \ie
$N_{_\textrm{R}}^{-2\ell} = N_{_\textrm{R}}^{-140}\sim \,10^{13}$ for a typical neutron
star.  Furthermore, the luminosity at these high modes should be strongly
suppressed as indicated by the factors ${\cal
  O}\left[\left(2\ell'-3\right)!!\right]^{2}$ appearing in the
denominator of the general expression for electromagnetic radiation
(\ref{em_radiation_toroidal}). This suppression, which is not a
general-relativistic effect but is present already in Newtonian gravity,
indicates that the pulsar effectively radiates energy at a vanishingly
small rate for these modes, which should be unlikely to be detected in
practice.

\subsection{Damping due to electromagnetic radiation from
spheroidal oscillations}

The electromagnetic power radiated when the star is subject to spheroidal
oscillations can be calculated in complete analogy to what done for
toroidal oscillations. An important difference, however, is that while
the electromagnetic radiation from toroidal oscillations is produced only
by perturbations in the radial component of the magnetic field,
spheroidal oscillations perturb all components of the magnetic field,
which therefore contribute to the emitted power. As an example, the power
radiated when the star is subject to dipolar and axisymmetric modes (\ie
for $\ell'=1,\, m'=0$) is given by
\begin{align}
\label{L_10_spheroid_eval}
    L_{_{\rm EM}}(1,0)
    &= \frac{c}{180\rm{\pi}}
    \left[\left(\frac{\eta_{_\textrm{R}}}{\omega_{_\textrm{R}} R}\right)
    \frac{h_{_\textrm{R}}}{N_{_\textrm{R}}}+
    \left(\frac{\xi_{_\textrm{R}}}{\omega_{_\textrm{R}} R}\right)
    \frac{f_{_\textrm{R}}}{N_{_\textrm{R}}}\right]^2
    \left({B_{_0}R}\right)^2
    \left(\frac{\omega_{_\textrm{R}}R}{c}\right)^6\cos^2\chi\
    \nonumber\\
    &\approx 5.3\times 10^{30}
    \left[\frac{\omega^2R^3}{GM}
    \frac{h_{_\textrm{R}}}{N^3_{_\textrm{R}}}+
    \frac{f_{_\textrm{R}}}{N^3_{_\textrm{R}}}\right]^2
    \left(\frac{\xi_{_\textrm{R}}}{\omega \textrm{R}}\right)
    \, B^2_{12}
    \, R^8_{6}
    \, \omega_3^{6} \cos^2\chi
    \quad {\rm erg\ s}^{-1} \,,
\end{align}
where and we have used the following relation between the radial
eigenfunctions [\cf equation (14.13) of \citet{Unno1989}]
\begin{equation}
\label{relation_eta_xi} \eta_{_\textrm{R}}=\xi_{_\textrm{R}}
         \frac{\omega^2R^3}{GM}\,.
\end{equation}
While the expression above is strictly valid at the surface of the star,
in the Cowling approximation and for a Newtonian star, it represents a
reasonable approximation and highlights that the energy losses for this
mode are comparable with the corresponding losses through toroidal
oscillations [\cf equation (\ref{em_radiation_toroidal})].

The ratio of the luminosities in the general-relativistic and
Newtonian approaches is thus given by
\begin{equation}
L_{_{\rm EM}}(\ell',m') / L_{_{\rm EM}}(\ell',m')\vert_{\rm Newt}
= \left(\frac{\omega^2R^3}{GM}\frac{h_{_\textrm{R}}}{N^3_{_\textrm{R}}}+
    \frac{f_{_\textrm{R}}}{N^3_{_\textrm{R}}}\right)
\left(\frac{\omega^2R^3}{2GM}+1\right)^{-1}\,.
\end{equation}
It should be noted that in contrast with the toroidal oscillations, the
electromagnetic energy losses through spheroidal modes is proportional to
the additional parameter $h_{_\textrm{R}}/N_{_\textrm{R}}$, which is responsible for
non-radial components of the surface magnetic field. However, since the
parameters $2h{_{_\textrm{R}}}$ and $f{_{_\textrm{R}}}$ are of the same order, and since
${\omega^2R^3}/2{GM}$ is $\sim 2.5\times 10^{-2}$ for a typical neutron
star with compactness $M/R=0.2$ oscillating at a frequency $\sim{\rm
  kHz}$, we can neglect the contribution coming from the pulsations in
the radial direction and write the amplification factor simply as
$\left({f_{_\textrm{R}}}/{N^3_{_\textrm{R}}}\right)^2$.

As an additional example we can consider the power emitted by a
purely quadrupolar mode (\ie with $\ell'=2, m'=2$) and express the
corresponding electromagnetic luminosity as
\begin{eqnarray}
\label{L_22_spheroid}
L_{_{\rm EM}}(2,2)
&\approx&
    7.2\times 10^{30}
    \left\{\left(\frac{\eta_{_\textrm{R}}}{\omega \textrm{R}}\right)^2
    \frac{5}{2}\left(\frac{5h_{_\textrm{R}}}{3N^4_{_\textrm{R}}}\right)^2
    +\left[\left(\frac{\xi_{_\textrm{R}}}{\omega R}\right)
    \frac{f_{_\textrm{R}}}{N^4_{_\textrm{R}}}-
    \left(\frac{\eta_{_\textrm{R}}}{\omega R}\right)
    \frac{2h_{_\textrm{R}}}{3N^4_{_\textrm{R}}}\right]^2\right\}
    \, B_{12}^2
    \, R_{6}^{10}
    \, \omega_3^{8} \cos^2\chi
    \quad {\rm erg\ s}^{-1} \,,
\end{eqnarray}
which is much smaller than the corresponding losses through toroidal
oscillations [\cf equation (\ref{L_22_toroid})]. This is because spheroidal
modes involve bulk compression and vertical motion, which have to do work
against the strong degeneracy pressure of the electrons in the outer
crust and the free neutrons in deeper layers and, of course, against the
strong vertical gravitational field.

Finally, in the limit $\omega_{_\textrm{R}} R \ll 1$, it is possible to derive the
following general expression for the electromagnetic power $L_{_{\rm
    EM}}(\ell',m')$ radiated by an arbitrary spheroidal $(\ell',m')$
oscillation mode with $\ell'>2$
\begin{eqnarray}
\label{em_radiation_spheroidal}
L_{_{\rm EM}}(\ell',m') = \frac{c}{32\rm{\pi}}
    \left[\frac{2\left(\ell'+1\right)f_{_\textrm{R}}\xi_{_\textrm{R}}-
    h_{_\textrm{R}}\eta_{_\textrm{R}}}{\left(\omega_{_\textrm{R}} R\right)N_{_\textrm{R}}}\right]^2
    B_{_0}^2R^2
    \left(\frac{\omega_{_\textrm{R}}R}{c}\right)^{2\ell^\prime}
     \frac{\left(\ell'-1\right)
    \left(\ell^{\prime 2}-m^{\prime 2}\right)}
    {\left[\left(2\ell'-3\right)!!\right]^2
        \ell'\left(2\ell'+1\right)\left(2\ell'-1\right)}
    \cos^2\chi \,.
\end{eqnarray}

\begin{table}
\label{spheroid_damping_time}
\caption{Damping times for spheroidal modes computed for NS13T8 model of
  a neutron star \citep{McDermott1988}. The different columns report
  characteristics of oscillation modes as frequency and kinetic energy,
  power of electromagnetic radiation, gravitational and electromagnetic
  damping times in the Newtonian and general-relativistic cases. The last
  column represents the ratio of the gravitational and electromagnetic
  time-scales. For the various nodes we use the standard notation in
  stellar seismology so that, for instance, the mode $_{l}p_n$ refers to
  a $p$ mode with harmonic index $\ell=l$ and overtone number $n$; for
  simplicity we limit ourselves to the first overtone. Furthermore, in
  the case of $g$ modes, the superscript '$s$' refers to the surface
  modes as these are the only $g$ modes that are relevant for our
  discussion.}
\begin{tabular}{|c|c|l|l|l|l|l|l|l|c}
\hline
 ${\rm Mode}$ & $\omega$  & $E_{_{\rm T}}$
 &  $L^{^{\rm Newt}}_{_{\rm EM}} $&
 $L^{^{\rm GR}}_{_{\rm EM}} $&
 $\tau_{_{\rm GW}}$ & $\tau^{^{\rm Newt}}_{_{\rm EM}}$ &
 $\tau^{^{\rm GR}}_{_{\rm EM}}$ &
 $\tau_{_{\rm GW}}$/$\tau^{^{\rm GR}}_{_{\rm EM}}$ & $\tau^{^{\rm Newt}}_{_{\rm EM}}/\tau^{^{\rm GR}}_{_{\rm EM}}$\\
 & (kHz) & (erg) & (erg/s) & (erg/s) & (s) & (s) & (s) & & \\
\hline
$_2p_{1}$ & $\phantom{0}75.6$ & $3.08\times 10^{51}$ & $1.44\times 10^{44}$
& $3.44\times 10^{44}$ & $3.80\times 10^{-4}$ & $4.3\times 10^{7}$
& $1.8\times 10^{7}$ & $2.1\times 10^{-11}$ & $\phantom{0}2.4$
\\
$_2f$ & $\phantom{0}28.6$ & $1.59\times 10^{52}$ & $2.38\times 10^{43}$ &
$7.41\times 10^{44}$ & $7.50\times 10^{-3}$ & $1.3\times 10^{9}$
& $4.3\times 10^{7}$ & $1.7\times 10^{-10}$ & $31.2$
\\
$_2s_{1}$ & $\phantom{00}8.6$ & $1.32\times 10^{54}$ & $5.13\times 10^{43}$
 & $1.12\times 10^{45}$ & $4.32\times 10^{4}$ &
$5.1\times 10^{10}$ & $2.4\times 10^{9}$ & $1.8\times 10^{-5}$
& $21.8$
\\
$_2i_{1}$ & $\phantom{00}0.3$ & $1.63\times 10^{53}$ & $5.49\times 10^{43}$ &
 $1.16\times 10^{45}$ & $8.64\times 10^{5}$ & $5.9\times 10^{9}$ &
 $2.8\times 10^{8}$ & $3.1\times 10^{-3}$ & $21.2$
\\
$_2g^s_{1}$ & $\phantom{0}0.14$ & $3.74\times 10^{41}$ & $5.49\times 10^{42}$
& $1.16\times 10^{45}$ & $2.90\times 10^{17}$ & $13\times 10^{-3}$
& $6.4\times 10^{-4}$ & $4.5\times 10^{20}$ & $20.3$
\\
  \hline
\end{tabular}
\end{table}

In Table~\ref{spheroid_damping_time}, we have reported the electromagnetic
damping times for some typical spheroidal modes and compared them with
the corresponding gravitational ones for $\ell\geq 2$. In addition, the
gravitational-radiation damping times $\tau_{_{\rm GW}}$ reported in
the fifth column were calculated by \citet{McDermott1988} using the general
expressions for the emission of gravitational radiation through
spheroidal oscillations given by \citet{Balbinski1982} and with a kinetic
energy
\begin{equation}
E_{_\textrm{T}} =
\frac{1}{2}\omega^2\int_0^R\rho\left[\tilde{\eta}^2+
\ell(\ell+1)\tilde{\xi}^2 \right]r^2\,{\rm d}r
\,.
\end{equation}
The results of our general-relativistic calculations are given in columns
4 and 7 of the Table, while the last two columns represent the ratio of
the gravitational and electromagnetic time-scales and the ratio of the
electromagnetic time-scales in the Newtonian and general-relativistic
case, respectively. A simple inspection of Table
\ref{spheroid_damping_time} is sufficient to reveal that gravitational
radiation is the most efficient damping mechanism for quadrupolar
spheroidal modes. This is obviously different with what has been reported
in Table~\ref{toroid_damping_time} for toroidal modes, but not
surprising. $g$ modes, in fact, are very inefficiently damped by
gravitational radiation, in spite of the fact that they involve bulk
motion of matter. This is mostly due to the frequencies of the $g$ modes,
that are comparatively low.

\subsection{Damping due to Joule heating}
\label{damping_joule}

All of the results so far have been obtained assuming that the stellar
interior has an infinite conductivity ($\sigma =\infty$). As discussed in
paper I, this is a very good approximation over the time-scale of a few
oscillations. Yet, for realistic stars, the electric conductivity will be
high but finite, and thus, the magnetic-field lines are not frozen
perfectly slipping slightly during oscillations. This motion of the
conducting matter with respect to the field lines generates electric
currents and Ohmic dissipation of these currents results in Joule heating
which will act electromagnetic damping mechanism for the stellar
interior.

To estimate the Joule heating we use the Maxwell equations, whose first
pair is given by
\begin{equation}
\label{maxwell_firstpair}
3! \,\partial_{[\gamma} F_{\alpha \beta]} =  2 \left(\partial_{\gamma} F_{\alpha \beta}
        + \partial_{\beta} F_{\gamma \alpha} + \partial_{\alpha} F_{\beta \gamma}
        \right) = 0 \,,
\end{equation}
where, for an observer with four-velocity $u^{\alpha}$, the covariant
components of the electromagnetic (Faraday) tensor $F_{\alpha\beta}$ are
given by
\begin{equation}
\label{fab_def}
F_{\alpha\beta} \equiv 2 u_{[\alpha} E_{\beta]} +
        \eta_{\alpha\beta\gamma\delta}u^\gamma B^\delta \,.
\end{equation}
Here $T_{[\alpha \beta]} \equiv \frac{1}{2}(T_{\alpha \beta} - T_{\beta
  \alpha})$ and $\eta_{\alpha\beta\gamma\delta}$ is totally antisymmetric
symbol employed in the definition of the Levi-Civita tensor
$\epsilon_{\alpha \beta \gamma \delta}$ \citep{Rezzolla_book:2013}
\begin{align}
&\eta_{\alpha\beta\gamma\delta}=-\frac{1}{\sqrt{-g}}
        \epsilon_{\alpha\beta\gamma\delta} \,,
&\eta^{\alpha\beta\gamma\delta}=
        \sqrt{-g}\epsilon^{\alpha\beta\gamma\delta} \,.
\end{align}
Similarly, the second pair of Maxwell equations is given by
\begin{equation}
\label{maxwell_secondpair}
\nabla_\beta F^{\alpha \beta} = 4 \rm{\pi} J^{\alpha} =
    4\rm{\pi} (\rho_e u^\alpha + j^\alpha) \,,
\end{equation}
where $u^\alpha$ is the fluid four-velocity, $\rho_e$ is the charge density
and $\boldsymbol{J}$ the {total} electric-charge current, such that
the divergence of electromagnetic energy-momentum tensor is
\begin{equation}
\label{diverg}
\nabla_\beta T^{\alpha\beta}=\frac{1}{4\rm{\pi}}\nabla_\beta \left(F^{\alpha\mu}
    F^{\beta}_{\;\mu}-\frac{1}{4}g^{\alpha\beta}
    F^{\mu\nu}F_{\mu\nu}\right)=F^{\mu\alpha}
    J_{\mu}\,.
\end{equation}
Note that the total electric-charge current includes the {conduction}
current $j^\alpha$, which is associated with electrons having electrical
conductivity $\sigma$, and that Ohm's law can be written as
\begin{equation}
\label{ohm}
j_\alpha := \sigma F_{ \alpha \beta}u^\beta \,.
\end{equation}
Contracting equation (\ref{diverg}) with the fluid four-velocity
$u_\alpha$ and using Ohm's law (\ref{ohm}), we obtain the change in
electromagnetic energy (or Joule dissipation rate) along the worldlines
of the conducting medium and hence
\begin{equation}
\label{joule}
Q_J=u_\alpha \nabla_\beta T^{\alpha\beta}=
    u_\alpha F^{\beta\alpha}
    J_{\beta} =\frac{1}{\sigma}j^2\,.
\end{equation}
This expression is quadratic in the current $j:= \left(j^\alpha
j_\alpha\right)^{1/2}$, linear in the resistance $1/\sigma$, and
coincides with expression (24) derived by \citet{Page2000} through an
alternative approach.

A generic misaligned dipolar magnetic field will be sustained by an
electric current with components $j^{\hat\alpha} := \left(0, j^{\hat r},
j^{\hat\theta}, j^{\hat\phi}\right)$~\footnote{Note that $J^{\hat r}=0$
  for toroidal oscillations.}  governed by the relevant components of the
Maxwell equations (\ref{maxwell_secondpair})
\begin{eqnarray}
\label{current_1}
&& J^{\hat r} = \frac{c}{4\rm{\pi} r}
    \left[\partial_{\theta}\left(\sin\theta B^{\hat\phi}\right) - \partial_{\phi}
        B^{\hat \theta}\right]        \,,
\\\nonumber\\
\label{current_2}
&& J^{\hat\theta} = \frac{c}{4\rm{\pi} r\sin\theta}
    \left[\partial_{\phi}B^{\hat r} - e^{-(\Phi+\Lambda)}
    \sin\theta \,\partial_{\textrm{r}}\left(re^{\Phi}
        B^{\hat \phi} \right)\right]        \,,
\\\nonumber\\
\label{current_3}
&& J^{\hat\phi} = \frac{c}{4\rm{\pi} r}
    \left[e^{-(\Phi+\Lambda)}\,\partial_{\textrm{r}}\left(e^{\Phi}r
    B^{\hat \theta} \right)
    -\partial_{\theta}B^{\hat r}\right] \,,
\end{eqnarray}
where $g_{00}=-e^{2\Phi}$ and $g_{11}=e^{-2\Lambda}$ are the components
of the metric tensor in the interior of the neutron star and solutions of
the Einstein equations.

Inserting the electric currents (\ref{current_1}) -- (\ref{current_3})
into (\ref{joule}), the resulting expression for the Joule dissipation
rate is
\begin{equation}
\label{joule_1}
Q_J=\frac{c^2}{16\rm{\pi}^2\sigma r^2}
\left\{\left[\partial_{\theta}\left(\sin\theta B^{\hat\phi}\right) - \partial_{\phi}
        B^{\hat \theta}\right]+\frac{1}{\sin^2\theta}
    \left[\partial_{\phi}B^{\hat r} - e^{-(\Phi+\Lambda)}
    \sin\theta \,\partial_{\textrm{r}}\left(re^{\Phi}
        B^{\hat \phi} \right)\right]^2+
    \left[e^{-(\Phi+\Lambda)}\,\partial_{\textrm{r}}\left(e^{\Phi}r
    B^{\hat \theta} \right)
    -\partial_{\theta}B^{\hat r}\right]^2\right\}   \,.
\end{equation}

Assuming that the stellar oscillations are toroidal, expanding the
magnetic field $B^{\hat\alpha}$ in a power series in terms of a
perturbation parameter ${\tilde\eta}/R$ (with ${\tilde\eta}/R\ll 1$), and
truncating at first order, \ie
\begin{equation}
B^{\hat\alpha}=B^{\hat\alpha}_{_0}
\left[1+\left(\frac{{\tilde\eta}}{R}\right)\right]\,,
\end{equation}
where $B^{\hat\alpha}_{_0}$ is the unperturbed magnetic field, the
expression for Joule heating (\ref{joule_1}) will take the form
\begin{equation}
\label{joule_1_{_0}_m}
Q_J=\frac{c^2e^{-2\Lambda}}{16\rm{\pi}^2\sigma}
        \left[\partial_{\rm r}\left(\frac{{\tilde\eta}}{R}\right)\right]^2
    \left[\left(B^{\hat\theta}_{0}\right)^2+
    \left(B^{\hat\phi}_{0}\right)^2
        \right]=
    e^{-2\Lambda}
        \left(\frac{{\tilde\eta}}{R}\right)^2\frac{1}{4\rm{\pi}}
    \left[\left(B^{\hat\theta}_{0}\right)^2+
    \left(B^{\hat\phi}_{0}\right)^2
        \right] \frac{c^2}{4\rm{\pi}\sigma R^2} \,,
\end{equation}
where we have used the Maxwell equations for the unperturbed magnetic
field
\begin{eqnarray}
&&e^{-(\Phi+\Lambda)}\,\partial_{r}\left(e^{\Phi}r
        B^{\hat \theta}_{_0} \right)
        -\partial_{\theta}B^{\hat r}_{0}=0 \,,
\\\nonumber\\
&&\partial_{\phi}B^{\hat r}_{0} - e^{-(\Phi+\Lambda)}
    \sin\theta \,\partial_{\textrm{r}}\left(re^{\Phi}
        B^{\hat \phi}_{_0} \right)=0 \,,
\end{eqnarray}
and have assumed $\partial_{\textrm{r}}{\tilde\eta}\sim{\tilde\eta}/R$ as by
\citet{McDermott1984}. Note that the Joule heating (\ref{joule_1_{_0}_m})
has two modifications with respect to its Newtonian equivalent. The first
one is via the metric correction $e^{-2\Lambda}$, while the second one is
through the general-relativistic amplification of the stellar magnetic
field [\cf equations (\ref{gr_fnctn}), (\ref{mf_1}) -- (\ref{mf_3}]. We
can now estimate an upper limit for the Joule dissipation by evaluating
expression (\ref{joule_1_{_0}_m}) at the surface of the star to obtain
\begin{equation}
\label{joule_estim}
L_J=\left(\frac{{\tilde\eta}}{R}\right)^2\frac{B_{_0}^2}{4\rm{\pi}}
    \left(2h_{_\textrm{R}}N_{_\textrm{R}}^2\right)^2
    \left(\frac{4\rm{\pi}}{3}R^3\right)
    \frac{c^2}{4\rm{\pi}\sigma R^2} \, ,
\end{equation}
where the quantity $\tau_{_\sigma}:= 4\rm{\pi}\sigma R^2/c^2$ is the
characteristic decay time-scale. Using an approximate expression for the
electrical conductivity given as \citep{Lamb:1991,Rezzolla01}
\begin{equation}
\sigma\approx 10^{23}\left(\frac{10^{8}\,{\rm K}}{T}\right)^{2}
\left(\frac{\rho}{10^{10}\,{\rm g\, cm^{-3}}}\right)^{3/4}\,{\rm
s^{-1}}\,,
\end{equation}
we obtain the time-scale of magnetic-field decay to be $\sim 10^{7} -
10^{8}\, {\rm yr}$. For a typical conductivity of the stellar crust
$\sigma\sim 10^{23}\,{\rm s^{-1}}$, and assuming the oscillations to have
the largest amplitude at the stellar surface, we can estimate the
magnitude of the energy loss due to electromagnetic heating as given by
\begin{equation}
\label{joule_estimation}
L_J\approx 0.23\times 10^{28}
    \left(\frac{{\tilde\eta}}{R}\right)^2
    \left(2h_{_\textrm{R}}N_{_\textrm{R}}^2\right)^2
    \left(\frac{B_{_0}}{10^{12}\,{\rm G}}\right)^2
    \left(\frac{R}{10^6 \,{\rm cm}}\right)
    \left(\frac{10^{23}\,{\rm s^{-1}}}{\sigma}\right)
    \quad {\rm erg\ s}^{-1}\,.
\end{equation}
Because of the intrinsically high electric conductivity, the Ohmic loss
rate~(\ref{joule_estimation}) for a canonical neutron star is several
orders of magnitude smaller than the radiation loss rates given by the
equation~(\ref{em_radiation_toroidal}) for dipolar and quadrupolar
toroidal modes and given by expressions (\ref{L_10_spheroid_eval}),
(\ref{L_22_spheroid}), (\ref{em_radiation_spheroidal}) for the spheroidal
modes. Furthermore, the general-relativistic contribution in the Ohmic
dissipation~(\ref{joule_estimation}) mainly arises through the change of
stellar magnetic field by the factor $2h_{_\textrm{R}}$.

In summary, the results presented in this section via
equations~(\ref{em_radiation_toroidal}),
(\ref{L_10_spheroid_eval})--(\ref{L_22_spheroid}) and
(\ref{joule_estimation}), show that there are general-relativistic
corrections both for the radiative electromagnetic losses produced by
toroidal/spheroidal modes and by the Joule heating in the case of finite
conductivity. However, while the latter can be reasonably ignored, there
are situations in which the electromagnetic energy damping could be
either comparable or larger than the energy losses via gravitational
radiation.

\section{Conclusion}
\label{conclusions}

In previous work of ours (paper I) we have developed a
general-relativistic formalism describing the vacuum electrodynamics of
an oscillating magnetized relativistic star. The assumptions needed to
make this problem tractable analytically are those of slow rotation,
infinite electric conductivity and that the electromagnetic energy does
not have a feedback on to the background gravitational field. In this
paper, we have applied the formalism to obtain explicit analytical
expressions for the electric and magnetic fields produced by the most
common modes of oscillation both in the vicinity of the star, where they
are quasi-stationary, and far away from it, where they behave as
electromagnetic waves. In this way, we have revisited and extended to a
general-relativistic context some of the work presented by
\citet{Muslimov1986} within Newtonian gravity.

In addition, we have considered the important issue in the
asteroseismology of compact and magnetized stars of determining the
dissipation mechanism which is most efficient in damping the
low-multipolarity oscillations. More specifically, we have computed the
electromagnetic radiation generated when a magnetized neutron star is
subject to either toroidal or spheroidal oscillations and computed the
energy losses in the form of Poynting fluxes, Joule heating and Ohmic
dissipation. This has allowed us to extend to a general-relativistic
context the classical Newtonian estimates of \citet{McDermott1988} for
the damping times of oscillating magnetized neutron stars.

In summary, despite the fact that a number of factors concur in
determining what is the main damping mechanism of the oscillations, \eg
the type of mode, the magnetic-field strength and the compactness of the
star, we have found that the following results to be robust for a typical
neutron star with a dipolar magnetic field of $\sim 10^{12} \,{\rm
  G}$. First, the general-relativistic corrections to the electromagnetic
fields lead to damping time-scales due to electromagnetic losses which are
at least one order of magnitude smaller than their Newtonian
counterparts. Secondly, $f$, $p$, $i$ and $s$ modes are suppressed more
efficiently by gravitational losses than by electromagnetic ones; the
only exception to this behaviour is given by $g$ modes, probably because
of the low typical frequencies of these modes. Finally, Joule heating is
not as an important damping mechanism in general relativity as it is in
Newtonian gravity.

The results obtained here could find at least two important
applications. First, through a more precise characterization of the
general-relativistic corrections, the electromagnetic waves emitted by an
oscillating star can be used to deduce, in conjunction with the
corresponding gravitational-wave signal, important constrains on the
properties of matter at nuclear density. Second, by better estimating the
energy loss to the emission of electromagnetic waves it is possible to
determine more accurately the time-scale over which oscillations of
different type can survive in a relativistic magnetized star.  For
example, the results presented here could be used to estimate the damping
of magnetar oscillation now that the oscillation eigenfrequencies and the
kinetic energy of magnetar oscillations are becoming increasingly more
accurate \citep{Cerda2009, Sotani2009, Gabler2011, Hambaryan2011,
  Gabler2014}.

\section*{Acknowledgements}

This research was partially supported by the Volkswagen Stiftung (Grant
86 866), by the LOEWE-Program in HIC for FAIR, by ``NewCompStar'', COST
Action MP1304, and by the European Union's Horizon 2020 Research and
Innovation Programme under grant agreement No. 671698 (call
FETHPC-1-2014, project ExaHyPE). BJA is also supported in part by the
project F2-FA-F113 of the UzAS and by the ICTP through the projects
OEA-NET-76, OEA-PRJ-29. BJA thanks the Institut f{\"u}r Theoretische
Physik for warm hospitality during his stay in Frankfurt.

\bibliographystyle{mnras.bst}

\begin{thebibliography}{}
\makeatletter
\relax
\def\mn@urlcharsother{\let\do\@makeother \do\$\do\&\do\#\do\^\do\_\do\%\do\~}
\def\mn@doi{\begingroup\mn@urlcharsother \@ifnextchar [ {\mn@doi@}
  {\mn@doi@[]}}
\def\mn@doi@[#1]#2{\def\@tempa{#1}\ifx\@tempa\@empty \href
  {http://dx.doi.org/#2} {doi:#2}\else \href {http://dx.doi.org/#2} {#1}\fi
  \endgroup}
\def\mn@eprint#1#2{\mn@eprint@#1:#2::\@nil}
\def\mn@eprint@arXiv#1{\href {http://arxiv.org/abs/#1} {{\tt arXiv:#1}}}
\def\mn@eprint@dblp#1{\href {http://dblp.uni-trier.de/rec/bibtex/#1.xml}
  {dblp:#1}}
\def\mn@eprint@#1:#2:#3:#4\@nil{\def\@tempa {#1}\def\@tempb {#2}\def\@tempc
  {#3}\ifx \@tempc \@empty \let \@tempc \@tempb \let \@tempb \@tempa \fi \ifx
  \@tempb \@empty \def\@tempb {arXiv}\fi \@ifundefined
  {mn@eprint@\@tempb}{\@tempb:\@tempc}{\expandafter \expandafter \csname
  mn@eprint@\@tempb\endcsname \expandafter{\@tempc}}}

\bibitem[\protect\citeauthoryear{{Abdikamalov}, {Ahmedov}  \&
  {Miller}}{{Abdikamalov} et~al.}{2009}]{Abdikamalov2009}
{Abdikamalov} E.~B.,  {Ahmedov} B.~J.,   {Miller} J.~C.,  2009, \mn@doi [Mon.
  Not. R. Astron. Soc.] {10.1111/j.1365-2966.2009.14540.x}, \href
  {http://adsabs.harvard.edu/abs/2009MNRAS.395..443A} {395, 443}

\bibitem[\protect\citeauthoryear{{Andersson}}{{Andersson}}{1998}]{Andersson1998}
{Andersson} N.,  1998, \mn@doi [Astrophys. J.] {10.1086/305919}, \href
  {http://adsabs.harvard.edu/abs/1998ApJ...502..708A} {502, 708}

\bibitem[\protect\citeauthoryear{{Arfken} \& {Weber}}{{Arfken} \&
  {Weber}}{2005}]{Arfken2005}
{Arfken} G.~B.,  {Weber} H.~J.,  2005, Mathematical methods for physicists, 6th
  ed..
Elsevier, Amsterdam

\bibitem[\protect\citeauthoryear{{Asai} \& {Lee}}{{Asai} \&
  {Lee}}{2014}]{Asai2014}
{Asai} H.,  {Lee} U.,  2014, \mn@doi [Astrophys. J.]
  {10.1088/0004-637X/790/1/66}, \href
  {http://adsabs.harvard.edu/abs/2014ApJ...790...66A} {790, 66}

\bibitem[\protect\citeauthoryear{{Balbinski} \& {Schutz}}{{Balbinski} \&
  {Schutz}}{1982}]{Balbinski1982}
{Balbinski} E.,  {Schutz} B.~F.,  1982, \mn@doi [Mon. Not. Roy. Astr. Soc.]
  {10.1093/mnras/200.1.43P}, \href
  {http://adsabs.harvard.edu/abs/1982MNRAS.200P..43B} {200, 43P}

\bibitem[\protect\citeauthoryear{{Cerd{\'a}-Dur{\'a}n}, {Stergioulas}  \&
  {Font}}{{Cerd{\'a}-Dur{\'a}n} et~al.}{2009}]{Cerda2009}
{Cerd{\'a}-Dur{\'a}n} P.,  {Stergioulas} N.,   {Font} J.~A.,  2009, \mn@doi
  [Mon. Not. R. Astron. Soc.] {10.1111/j.1365-2966.2009.15056.x}, \href
  {http://adsabs.harvard.edu/abs/2009MNRAS.397.1607C} {397, 1607}

\bibitem[\protect\citeauthoryear{{Chugunov} \& {Yakovlev}}{{Chugunov} \&
  {Yakovlev}}{2005}]{Chugunov2005}
{Chugunov} A.~I.,  {Yakovlev} D.~G.,  2005, \mn@doi [Astronomy Reports]
  {10.1134/1.2045323}, \href
  {http://adsabs.harvard.edu/abs/2005ARep...49..724C} {49, 724}

\bibitem[\protect\citeauthoryear{{Ciolfi} \& {Rezzolla}}{{Ciolfi} \&
  {Rezzolla}}{2012}]{Ciolfi2012}
{Ciolfi} R.,  {Rezzolla} L.,  2012, \mn@doi [Astrophys. J.]
  {10.1088/0004-637X/760/1/1}, \href
  {http://adsabs.harvard.edu/abs/2012ApJ...760....1C} {760, 1}

\bibitem[\protect\citeauthoryear{{Ciolfi}, {Lander}, {Manca}  \&
  {Rezzolla}}{{Ciolfi} et~al.}{2011}]{Ciolfi2011}
{Ciolfi} R.,  {Lander} S.~K.,  {Manca} G.~M.,   {Rezzolla} L.,  2011, \mn@doi
  [Astrophys. J.] {10.1088/2041-8205/736/1/L6}, \href
  {http://adsabs.harvard.edu/abs/2011ApJ...736L...6C} {736, L6}

\bibitem[\protect\citeauthoryear{{Clemens} \& {Rosen}}{{Clemens} \&
  {Rosen}}{2004}]{Clemens2004}
{Clemens} J.~C.,  {Rosen} R.,  2004, \mn@doi [Astrophys. J.] {10.1086/421013},
  \href {http://adsabs.harvard.edu/abs/2004ApJ...609..340C} {609, 340}

\bibitem[\protect\citeauthoryear{{Clemens} \& {Rosen}}{{Clemens} \&
  {Rosen}}{2008}]{Clemens2008}
{Clemens} J.~C.,  {Rosen} R.,  2008, \mn@doi [Astrophys. J.] {10.1086/587474},
  \href {http://adsabs.harvard.edu/abs/2008ApJ...680..664C} {680, 664}

\bibitem[\protect\citeauthoryear{{Colaiuda}, {Beyer}  \& {Kokkotas}}{{Colaiuda}
  et~al.}{2009}]{Colaiuda2009}
{Colaiuda} A.,  {Beyer} H.,   {Kokkotas} K.~D.,  2009, \mn@doi [Mon. Not. R.
  Astron. Soc.] {10.1111/j.1365-2966.2009.14878.x}, \href
  {http://adsabs.harvard.edu/abs/2009MNRAS.396.1441C} {396, 1441}

\bibitem[\protect\citeauthoryear{{Cuofano} \& {Drago}}{{Cuofano} \&
  {Drago}}{2010}]{Cuofano2010}
{Cuofano} C.,  {Drago} A.,  2010, \mn@doi [Phys. Rev. D]
  {10.1103/PhysRevD.82.084027}, \href
  {http://adsabs.harvard.edu/abs/2010PhRvD..82h4027C} {82, 084027}

\bibitem[\protect\citeauthoryear{{Cuofano}, {Dall'Osso}, {Drago}  \&
  {Stella}}{{Cuofano} et~al.}{2012}]{Cuofano2012}
{Cuofano} C.,  {Dall'Osso} S.,  {Drago} A.,   {Stella} L.,  2012, \mn@doi
  [Phys. Rev. D] {10.1103/PhysRevD.86.044004}, \href
  {http://adsabs.harvard.edu/abs/2012PhRvD..86d4004C} {86, 044004}

\bibitem[\protect\citeauthoryear{{Duncan}}{{Duncan}}{1998}]{Duncan1998}
{Duncan} R.~C.,  1998, \mn@doi [Astrophys. J. Lett.] {10.1086/311303}, \href
  {http://adsabs.harvard.edu/abs/1998ApJ...498L..45D} {498, L45+}

\bibitem[\protect\citeauthoryear{{El-Mezeini} \& {Ibrahim}}{{El-Mezeini} \&
  {Ibrahim}}{2010}]{El-Mezeini2010}
{El-Mezeini} A.~M.,  {Ibrahim} A.~I.,  2010, \mn@doi [Astrophys. J. Lett.]
  {10.1088/2041-8205/721/2/L121}, \href
  {http://adsabs.harvard.edu/abs/2010ApJ...721L.121E} {721, L121}

\bibitem[\protect\citeauthoryear{{Finn}}{{Finn}}{1990}]{Finn1990}
{Finn} L.~S.,  1990, Mon. Not. R. Astron. Soc., \href
  {http://adsabs.harvard.edu/abs/1990MNRAS.245...82F} {245, 82}

\bibitem[\protect\citeauthoryear{{Friedman} \& {Morsink}}{{Friedman} \&
  {Morsink}}{1998}]{Friedman1998}
{Friedman} J.~L.,  {Morsink} S.~M.,  1998, \mn@doi [Astrophys.J.]
  {10.1086/305920}, \href {http://adsabs.harvard.edu/abs/1998ApJ...502..714F}
  {502, 714}

\bibitem[\protect\citeauthoryear{Friedman, Lindblom  \& Lockitch}{Friedman
  et~al.}{2016}]{Friedman2015}
Friedman J.~L.,  Lindblom L.,   Lockitch K.~H.,  2016, \mn@doi [Phys. Rev. D]
  {10.1103/PhysRevD.93.024023}, 93, 024023

\bibitem[\protect\citeauthoryear{{Gabler}, {Cerd{\'a} Dur{\'a}n}, {Font},
  {M{\"u}ller}  \& {Stergioulas}}{{Gabler} et~al.}{2011}]{Gabler2011}
{Gabler} M.,  {Cerd{\'a} Dur{\'a}n} P.,  {Font} J.~A.,  {M{\"u}ller} E.,
  {Stergioulas} N.,  2011, \mn@doi [Mon. Not. R. Astron. Soc.]
  {10.1111/j.1745-3933.2010.00974.x}, \href
  {http://adsabs.harvard.edu/abs/2011MNRAS.410L..37G} {410, L37}

\bibitem[\protect\citeauthoryear{{Gabler}, {Cerd{\'a}-Dur{\'a}n},
  {Stergioulas}, {Font}  \& {M{\"u}ller}}{{Gabler} et~al.}{2012}]{Gabler2012}
{Gabler} M.,  {Cerd{\'a}-Dur{\'a}n} P.,  {Stergioulas} N.,  {Font} J.~A.,
  {M{\"u}ller} E.,  2012, \mn@doi [Mon. Not. R. Astron. Soc.]
  {10.1111/j.1365-2966.2012.20454.x}, \href
  {http://adsabs.harvard.edu/abs/2012MNRAS.421.2054G} {421, 2054}

\bibitem[\protect\citeauthoryear{{Gabler}, {Cerd{\'a}-Dur{\'a}n}, {Font},
  {M{\"u}ller}  \& {Stergioulas}}{{Gabler} et~al.}{2013}]{Gabler2013b}
{Gabler} M.,  {Cerd{\'a}-Dur{\'a}n} P.,  {Font} J.~A.,  {M{\"u}ller} E.,
  {Stergioulas} N.,  2013, \mn@doi [Mon. Not. R. Astron. Soc.]
  {10.1093/mnras/sts721}, \href
  {http://adsabs.harvard.edu/abs/2013MNRAS.430.1811G} {430, 1811}

\bibitem[\protect\citeauthoryear{{Gabler}, {Cerd{\'a}-Dur{\'a}n},
  {Stergioulas}, {Font}  \& {M{\"u}ller}}{{Gabler} et~al.}{2014}]{Gabler2014}
{Gabler} M.,  {Cerd{\'a}-Dur{\'a}n} P.,  {Stergioulas} N.,  {Font} J.~A.,
  {M{\"u}ller} E.,  2014, \mn@doi [Mon. Not. R. Astron. Soc.]
  {10.1093/mnras/stu1263}, \href
  {http://adsabs.harvard.edu/abs/2014MNRAS.443.1416G} {443, 1416}

\bibitem[\protect\citeauthoryear{{Gaertig} \& {Kokkotas}}{{Gaertig} \&
  {Kokkotas}}{2008}]{Gaertig:2008a}
{Gaertig} E.,  {Kokkotas} K.~D.,  2008, \mn@doi [Phys. Rev. D]
  {10.1103/PhysRevD.78.064063}, \href
  {http://adsabs.harvard.edu/abs/2008PhRvD..78f4063G} {78, 064063}

\bibitem[\protect\citeauthoryear{{Gaertig} \& {Kokkotas}}{{Gaertig} \&
  {Kokkotas}}{2011}]{Gaertig2011}
{Gaertig} E.,  {Kokkotas} K.~D.,  2011, \mn@doi [Phys. Rev. D]
  {10.1103/PhysRevD.83.064031}, \href
  {http://adsabs.harvard.edu/abs/2011PhRvD..83f4031G} {83, 064031}

\bibitem[\protect\citeauthoryear{{Glampedakis}, {Samuelsson}  \&
  {Andersson}}{{Glampedakis} et~al.}{2006}]{Glampedakis2006}
{Glampedakis} K.,  {Samuelsson} L.,   {Andersson} N.,  2006, \mn@doi [Mon. Not.
  R. Astron. Soc.] {10.1111/j.1745-3933.2006.00211.x}, \href
  {http://adsabs.harvard.edu/abs/2006MNRAS.371L..74G} {371, L74}

\bibitem[\protect\citeauthoryear{{Hambaryan}, {Neuh{\"a}user}  \&
  {Kokkotas}}{{Hambaryan} et~al.}{2011}]{Hambaryan2011}
{Hambaryan} V.,  {Neuh{\"a}user} R.,   {Kokkotas} K.~D.,  2011, \mn@doi
  [Astron. Astrophys.] {10.1051/0004-6361/201015273}, \href
  {http://adsabs.harvard.edu/abs/2011A%26A...528A..45H} {528, A45}

\bibitem[\protect\citeauthoryear{{Hansen} \& {Cioffi}}{{Hansen} \&
  {Cioffi}}{1980}]{Hansen1980}
{Hansen} C.~J.,  {Cioffi} D.~F.,  1980, \mn@doi [Astrophys.J.]
  {10.1086/158031}, \href {http://adsabs.harvard.edu/abs/1980ApJ...238..740H}
  {238, 740}

\bibitem[\protect\citeauthoryear{{Ho} \& {Lai}}{{Ho} \& {Lai}}{2000}]{Ho2000}
{Ho} W.~C.~G.,  {Lai} D.,  2000, \mn@doi [Astrophys. J.] {10.1086/317085},
  \href {http://adsabs.harvard.edu/abs/2000ApJ...543..386H} {543, 386}

\bibitem[\protect\citeauthoryear{{Huppenkothen}, {Heil}, {Watts}  \& {G{\"o}{\u
  g}{\"u}{\c s}}}{{Huppenkothen} et~al.}{2014}]{Huppenkothen2014a}
{Huppenkothen} D.,  {Heil} L.~M.,  {Watts} A.~L.,   {G{\"o}{\u g}{\"u}{\c s}}
  E.,  2014, \mn@doi [Astrophys. J.] {10.1088/0004-637X/795/2/114}, \href
  {http://adsabs.harvard.edu/abs/2014ApJ...795..114H} {795, 114}

\bibitem[\protect\citeauthoryear{{Israel} et~al.,}{{Israel}
  et~al.}{2005}]{Israel2005}
{Israel} G.~L.,  et~al., 2005, \mn@doi [Astrophys. J. Lett.] {10.1086/432615},
  \href {http://adsabs.harvard.edu/abs/2005ApJ...628L..53I} {628, L53}

\bibitem[\protect\citeauthoryear{{Kojima} \& {Okita}}{{Kojima} \&
  {Okita}}{2004}]{Kojima2004b}
{Kojima} Y.,  {Okita} T.,  2004, \mn@doi [Astrophys.J.] {10.1086/423706}, \href
  {http://adsabs.harvard.edu/abs/2004ApJ...614..922K} {614, 922}

\bibitem[\protect\citeauthoryear{{Kokkotas} \& {Schmidt}}{{Kokkotas} \&
  {Schmidt}}{1999}]{Kokkotas99a}
{Kokkotas} K.,  {Schmidt} B.,  1999, \mn@doi [Living Rev. Relativity]
  {10.12942/lrr-1999-2}, \href
  {http://adsabs.harvard.edu/abs/1999LRR.....2....2K} {2, 2}

\bibitem[\protect\citeauthoryear{{Konno} \& {Kojima}}{{Konno} \&
  {Kojima}}{2000}]{Konno2000b}
{Konno} K.,  {Kojima} Y.,  2000, \mn@doi [Progress of Theoretical Physics]
  {10.1143/PTP.104.1117}, \href
  {http://adsabs.harvard.edu/abs/2000PThPh.104.1117K} {104, 1117}

\bibitem[\protect\citeauthoryear{{Konno}, {Obata}  \& {Kojima}}{{Konno}
  et~al.}{1999}]{Konno1999}
{Konno} K.,  {Obata} T.,   {Kojima} Y.,  1999, Astron. Astrophys., \href
  {http://adsabs.harvard.edu/abs/1999A%26A...352..211K} {352, 211}

\bibitem[\protect\citeauthoryear{{Konno}, {Obata}  \& {Kojima}}{{Konno}
  et~al.}{2000}]{Konno2000a}
{Konno} K.,  {Obata} T.,   {Kojima} Y.,  2000, Astron. Astrophys., \href
  {http://adsabs.harvard.edu/abs/2000A%26A...356..234K} {356, 234}

\bibitem[\protect\citeauthoryear{{Lamb}}{{Lamb}}{1991}]{Lamb:1991}
{Lamb} F.~K.,  1991, in {Lambert} D.~L.,  ed.,  Astronomical Society of the
  Pacific Conference Series Vol. 20, Frontiers of Stellar Evolution. pp
  299--388

\bibitem[\protect\citeauthoryear{{Lasky}, {Zink}, {Kokkotas}  \&
  {Glampedakis}}{{Lasky} et~al.}{2011}]{Lasky2011}
{Lasky} P.~D.,  {Zink} B.,  {Kokkotas} K.~D.,   {Glampedakis} K.,  2011,
  \mn@doi [Astrophys. J.] {10.1088/2041-8205/735/1/L20}, \href
  {http://adsabs.harvard.edu/abs/2011ApJ...735L..20L} {735, L20}

\bibitem[\protect\citeauthoryear{{Lee}}{{Lee}}{2007}]{Lee2007}
{Lee} U.,  2007, \mn@doi [Mon. Not. R. Astron. Soc.]
  {10.1111/j.1365-2966.2006.11214.x}, \href
  {http://adsabs.harvard.edu/abs/2007MNRAS.374.1015L} {374, 1015}

\bibitem[\protect\citeauthoryear{{Levin}}{{Levin}}{2006}]{Levin2006}
{Levin} Y.,  2006, \mn@doi [Mon. Not. R. Astron. Soc.]
  {10.1111/j.1745-3933.2006.00155.x}, \href
  {http://adsabs.harvard.edu/abs/2006MNRAS.368L..35L} {368, L35}

\bibitem[\protect\citeauthoryear{{Levin}}{{Levin}}{2007}]{Levin2007}
{Levin} Y.,  2007, \mn@doi [Mon. Not. R. Astron. Soc.]
  {10.1111/j.1365-2966.2007.11582.x}, \href
  {http://adsabs.harvard.edu/abs/2007MNRAS.377..159L} {377, 159}

\bibitem[\protect\citeauthoryear{{Lin}, {Xu}  \& {Zhang}}{{Lin}
  et~al.}{2015}]{Lin2015}
{Lin} M.-X.,  {Xu} R.-X.,   {Zhang} B.,  2015, \mn@doi [Astrophys. J.]
  {10.1088/0004-637X/799/2/152}, \href
  {http://adsabs.harvard.edu/abs/2015ApJ...799..152L} {799, 152}

\bibitem[\protect\citeauthoryear{{McDermott}, {Savedoff}, {van Horn}, {Zweibel}
   \& {Hansen}}{{McDermott} et~al.}{1984}]{McDermott1984}
{McDermott} P.~N.,  {Savedoff} M.~P.,  {van Horn} H.~M.,  {Zweibel} E.~G.,
  {Hansen} C.~J.,  1984, \mn@doi [Astrophys. J.] {10.1086/162152}, \href
  {http://adsabs.harvard.edu/abs/1984ApJ...281..746M} {281, 746}

\bibitem[\protect\citeauthoryear{{McDermott}, {van Horn}, {Hansen}  \&
  {Buland}}{{McDermott} et~al.}{1985}]{McDermott1985}
{McDermott} P.~N.,  {van Horn} H.~M.,  {Hansen} C.~J.,   {Buland} R.,  1985,
  \mn@doi [Astrophys. J. Lett.] {10.1086/184553}, \href
  {http://adsabs.harvard.edu/abs/1985ApJ...297L..37M} {297, L37}

\bibitem[\protect\citeauthoryear{{McDermott}, {van Horn}  \&
  {Hansen}}{{McDermott} et~al.}{1988}]{McDermott1988}
{McDermott} P.~N.,  {van Horn} H.~M.,   {Hansen} C.~J.,  1988, \mn@doi
  [Astrophys. J.] {10.1086/166044}, \href
  {http://adsabs.harvard.edu/abs/1988ApJ...325..725M} {325, 725}

\bibitem[\protect\citeauthoryear{{Messios}, {Papadopoulos}  \&
  {Stergioulas}}{{Messios} et~al.}{2001}]{Messios2001}
{Messios} N.,  {Papadopoulos} D.~B.,   {Stergioulas} N.,  2001, \mn@doi [Mon.
  Not. R. Astron. Soc.] {10.1046/j.1365-8711.2001.04645.x}, \href
  {http://adsabs.harvard.edu/abs/2001MNRAS.328.1161M} {328, 1161}

\bibitem[\protect\citeauthoryear{{Morozova}, {Ahmedov}  \&
  {Zanotti}}{{Morozova} et~al.}{2010}]{Morozova2010}
{Morozova} V.~S.,  {Ahmedov} B.~J.,   {Zanotti} O.,  2010, \mn@doi [Mon. Not.
  R. Astron. Soc.] {10.1111/j.1365-2966.2010.17131.x}, \href
  {http://adsabs.harvard.edu/abs/2010MNRAS.408..490M} {408, 490}

\bibitem[\protect\citeauthoryear{{Morozova}, {Ahmedov}  \&
  {Zanotti}}{{Morozova} et~al.}{2012}]{Morozova2012}
{Morozova} V.~S.,  {Ahmedov} B.~J.,   {Zanotti} O.,  2012, \mn@doi [Mon. Not.
  R. Astron. Soc.] {10.1111/j.1365-2966.2011.19866.x}, \href
  {http://adsabs.harvard.edu/abs/2012MNRAS.419.2147M} {419, 2147}

\bibitem[\protect\citeauthoryear{{Morozova}, {Ahmedov}  \&
  {Zanotti}}{{Morozova} et~al.}{2014}]{Morozova2014}
{Morozova} V.~S.,  {Ahmedov} B.~J.,   {Zanotti} O.,  2014, \mn@doi [Mon. Not.
  R. Astron. Soc.] {10.1093/mnras/stu1486}, \href
  {http://adsabs.harvard.edu/abs/2014MNRAS.444.1144M} {444, 1144}

\bibitem[\protect\citeauthoryear{{Muslimov} \& {Harding}}{{Muslimov} \&
  {Harding}}{1997}]{Muslimov1997}
{Muslimov} A.,  {Harding} A.~K.,  1997, \mn@doi [Astrophys. J.]
  {10.1086/304457}, \href {http://adsabs.harvard.edu/abs/1997ApJ...485..735M}
  {485, 735}

\bibitem[\protect\citeauthoryear{{Muslimov} \& {Tsygan}}{{Muslimov} \&
  {Tsygan}}{1986}]{Muslimov1986}
{Muslimov} A.~G.,  {Tsygan} A.~I.,  1986, \mn@doi [Ap\&SS]
  {10.1007/BF00653898}, \href
  {http://adsabs.harvard.edu/abs/1986Ap%26SS.120...27M} {120, 27}

\bibitem[\protect\citeauthoryear{{Page}, {Geppert}  \& {Zannias}}{{Page}
  et~al.}{2000}]{Page2000}
{Page} D.,  {Geppert} U.,   {Zannias} T.,  2000, Astron. Astrophys., \href
  {http://adsabs.harvard.edu/abs/2000A%26A...360.1052P} {360, 1052}

\bibitem[\protect\citeauthoryear{{P{\'e}tri}}{{P{\'e}tri}}{2013}]{Petri2013}
{P{\'e}tri} J.,  2013, \mn@doi [Mon. Not. R. Astron. Soc.]
  {10.1093/mnras/stt798}, \href
  {http://adsabs.harvard.edu/abs/2013MNRAS.433..986P} {433, 986}

\bibitem[\protect\citeauthoryear{{Piro}}{{Piro}}{2005}]{Piro2005}
{Piro} A.~L.,  2005, \mn@doi [Astrophys. J. Lett.] {10.1086/499049}, \href
  {http://adsabs.harvard.edu/abs/2005ApJ...634L.153P} {634, L153}

\bibitem[\protect\citeauthoryear{{Rezzolla} \& {Ahmedov}}{{Rezzolla} \&
  {Ahmedov}}{2004}]{Rezzolla2004}
{Rezzolla} L.,  {Ahmedov} B.~J.,  2004, \mn@doi [Mon. Not. R. Astron. Soc.]
  {10.1111/j.1365-2966.2004.08006.x}, \href
  {http://adsabs.harvard.edu/abs/2004MNRAS.352.1161R} {352, 1161}

\bibitem[\protect\citeauthoryear{Rezzolla \& Zanotti}{Rezzolla \&
  Zanotti}{2001}]{Rezzolla01}
Rezzolla L.,  Zanotti O.,  2001, Journ. of Fluid Mech., 449, 395

\bibitem[\protect\citeauthoryear{{Rezzolla} \& {Zanotti}}{{Rezzolla} \&
  {Zanotti}}{2013}]{Rezzolla_book:2013}
{Rezzolla} L.,  {Zanotti} O.,  2013, {Relativistic Hydrodynamics}.
Oxford University Press, Oxford, UK,
  \mn@doi{10.1093/acprof:oso/9780198528906.001.0001}

\bibitem[\protect\citeauthoryear{{Rezzolla}, {Lamb}  \& {Shapiro}}{{Rezzolla}
  et~al.}{2000}]{Rezzolla00}
{Rezzolla} L.,  {Lamb} F.~K.,   {Shapiro} S.~L.,  2000, \mn@doi [Astrophys. J.
  Lett.] {10.1086/312539}, \href
  {http://adsabs.harvard.edu/abs/2000ApJ...531L.139R} {531, L139}

\bibitem[\protect\citeauthoryear{{Rezzolla}, {Ahmedov}  \& {Miller}}{{Rezzolla}
  et~al.}{2001a}]{Rezzolla2001b}
{Rezzolla} L.,  {Ahmedov} B.~J.,   {Miller} J.~C.,  2001a, Found. Phys., \href
  {http://adsabs.harvard.edu/abs/2001gr.qc.....8057R} {31, 1051}

\bibitem[\protect\citeauthoryear{{Rezzolla}, {Ahmedov}  \& {Miller}}{{Rezzolla}
  et~al.}{2001b}]{Rezzolla2001}
{Rezzolla} L.,  {Ahmedov} B.~J.,   {Miller} J.~C.,  2001b, \mn@doi [Mon. Not.
  R. Astron. Soc.] {10.1046/j.1365-8711.2001.04161.x}, \href
  {http://adsabs.harvard.edu/abs/2001MNRAS.322..723R} {322, 723}

\bibitem[\protect\citeauthoryear{{Richardson}, {van Horn}, {Ratcliff}  \&
  {Malone}}{{Richardson} et~al.}{1982}]{Richardson1982}
{Richardson} M.~B.,  {van Horn} H.~M.,  {Ratcliff} K.~F.,   {Malone} R.~C.,
  1982, \mn@doi [Astrophys. J.] {10.1086/159865}, \href
  {http://adsabs.harvard.edu/abs/1982ApJ...255..624R} {255, 624}

\bibitem[\protect\citeauthoryear{Rose}{Rose}{1955}]{Rose1955}
Rose M.~E.,  1955, Multipole Fields.
John Wiley, New York

\bibitem[\protect\citeauthoryear{{Rosen} \& {Clemens}}{{Rosen} \&
  {Clemens}}{2008}]{Rosen2008}
{Rosen} R.,  {Clemens} J.~C.,  2008, \mn@doi [Astrophys. J.] {10.1086/587476},
  \href {http://adsabs.harvard.edu/abs/2008ApJ...680..671R} {680, 671}

\bibitem[\protect\citeauthoryear{{S{\'a}}}{{S{\'a}}}{2004}]{Sa2004}
{S{\'a}} P.~M.,  2004, \mn@doi [Phys. Rev. D] {10.1103/PhysRevD.69.084001},
  \href {http://adsabs.harvard.edu/abs/2004PhRvD..69h4001S} {69, 084001}

\bibitem[\protect\citeauthoryear{{S{\'a}} \& {Tom{\'e}}}{{S{\'a}} \&
  {Tom{\'e}}}{2006}]{Sa2006}
{S{\'a}} P.~M.,  {Tom{\'e}} B.,  2006, \mn@doi [Phys. Rev. D]
  {10.1103/PhysRevD.74.044011}, \href
  {http://adsabs.harvard.edu/abs/2006PhRvD..74d4011S} {74, 044011}

\bibitem[\protect\citeauthoryear{{Samuelsson} \& {Andersson}}{{Samuelsson} \&
  {Andersson}}{2007}]{Samuelsson2007}
{Samuelsson} L.,  {Andersson} N.,  2007, \mn@doi [Mon. Not. R. Astron. Soc.]
  {10.1111/j.1365-2966.2006.11147.x}, \href
  {http://adsabs.harvard.edu/abs/2007MNRAS.374..256S} {374, 256}

\bibitem[\protect\citeauthoryear{{Schumaker} \& {Thorne}}{{Schumaker} \&
  {Thorne}}{1983}]{Schumaker1983}
{Schumaker} B.~L.,  {Thorne} K.~S.,  1983, Mon. Not. R. Astron. Soc., \href
  {http://adsabs.harvard.edu/abs/1983MNRAS.203..457S} {203, 457}

\bibitem[\protect\citeauthoryear{{Sotani} \& {Kokkotas}}{{Sotani} \&
  {Kokkotas}}{2009}]{Sotani2009}
{Sotani} H.,  {Kokkotas} K.~D.,  2009, \mn@doi [Mon. Not. R. Astron. Soc.]
  {10.1111/j.1365-2966.2009.14631.x}, \href
  {http://adsabs.harvard.edu/abs/2009MNRAS.395.1163S} {395, 1163}

\bibitem[\protect\citeauthoryear{{Sotani}, {Kokkotas}  \&
  {Stergioulas}}{{Sotani} et~al.}{2007}]{Sotani2007}
{Sotani} H.,  {Kokkotas} K.~D.,   {Stergioulas} N.,  2007, \mn@doi [Mon. Not.
  Roy. Astron. Soc.] {10.1111/j.1365-2966.2006.11304.x}, \href
  {http://adsabs.harvard.edu/abs/2007MNRAS.375..261S} {375, 261}

\bibitem[\protect\citeauthoryear{{Sotani}, {Kokkotas}  \&
  {Stergioulas}}{{Sotani} et~al.}{2008}]{Sotani2008}
{Sotani} H.,  {Kokkotas} K.~D.,   {Stergioulas} N.,  2008, \mn@doi [Mon. Not.
  R. Astron. Soc.] {10.1111/j.1745-3933.2007.00420.x}, \href
  {http://adsabs.harvard.edu/abs/2008MNRAS.385L...5S} {385, L5}

\bibitem[\protect\citeauthoryear{{Strohmayer} \& {Watts}}{{Strohmayer} \&
  {Watts}}{2005}]{Strohmayer2005}
{Strohmayer} T.~E.,  {Watts} A.~L.,  2005, \mn@doi [ApJL] {10.1086/497911},
  \href {http://adsabs.harvard.edu/abs/2005ApJ...632L.111S} {632, L111}

\bibitem[\protect\citeauthoryear{{Strohmayer} \& {Watts}}{{Strohmayer} \&
  {Watts}}{2006}]{Strohmayer2006}
{Strohmayer} T.~E.,  {Watts} A.~L.,  2006, \mn@doi [Astrophys. J]
  {10.1086/508703}, \href {http://adsabs.harvard.edu/abs/2006ApJ...653..593S}
  {653, 593}

\bibitem[\protect\citeauthoryear{{Timokhin}, {Bisnovatyi-Kogan}  \&
  {Spruit}}{{Timokhin} et~al.}{2000}]{Timokhin2000}
{Timokhin} A.~N.,  {Bisnovatyi-Kogan} G.~S.,   {Spruit} H.~C.,  2000, \mn@doi
  [Mon. Not. R. Astron. Soc.] {10.1046/j.1365-8711.2000.03535.x}, \href
  {http://adsabs.harvard.edu/abs/2000MNRAS.316..734T} {316, 734}

\bibitem[\protect\citeauthoryear{{Turolla}, {Zane}  \& {Watts}}{{Turolla}
  et~al.}{2015}]{Turolla2015}
{Turolla} R.,  {Zane} S.,   {Watts} A.~L.,  2015, \mn@doi [Reports on Progress
  in Physics] {10.1088/0034-4885/78/11/116901}, \href
  {http://adsabs.harvard.edu/abs/2015RPPh...78k6901T} {78, 116901}

\bibitem[\protect\citeauthoryear{{Unno}, {Osaki}, {Ando}, {Saio}  \&
  {Shibahashi}}{{Unno} et~al.}{1989}]{Unno1989}
{Unno} W.,  {Osaki} Y.,  {Ando} H.,  {Saio} H.,   {Shibahashi} H.,  1989,
  {Nonradial oscillations of stars}.
University of Tokyo Press

\bibitem[\protect\citeauthoryear{{Vavoulidis}, {Stavridis}, {Kokkotas}  \&
  {Beyer}}{{Vavoulidis} et~al.}{2007}]{Vavoulidis2007}
{Vavoulidis} M.,  {Stavridis} A.,  {Kokkotas} K.~D.,   {Beyer} H.,  2007,
  \mn@doi [Mon. Not. R. Astron. Soc.] {10.1111/j.1365-2966.2007.11706.x}, \href
  {http://adsabs.harvard.edu/abs/2007MNRAS.377.1553V} {377, 1553}

\bibitem[\protect\citeauthoryear{{Watts} \& {Strohmayer}}{{Watts} \&
  {Strohmayer}}{2006}]{Watts2006}
{Watts} A.~L.,  {Strohmayer} T.~E.,  2006, \mn@doi [Astrophys. J.]
  {10.1086/500735}, \href {http://cdsads.u-strasbg.fr/abs/2006ApJ...637L.117W}
  {637, L117}

\bibitem[\protect\citeauthoryear{{Watts} \& {Strohmayer}}{{Watts} \&
  {Strohmayer}}{2007a}]{Watts07b}
{Watts} A.~L.,  {Strohmayer} T.~E.,  2007a, \mn@doi [Advances in Space
  Research] {10.1016/j.asr.2006.12.021}, \href
  {http://adsabs.harvard.edu/abs/2007AdSpR..40.1446W} {40, 1446}

\bibitem[\protect\citeauthoryear{{Watts} \& {Strohmayer}}{{Watts} \&
  {Strohmayer}}{2007b}]{Watts07a}
{Watts} A.~L.,  {Strohmayer} T.~E.,  2007b, \mn@doi [Ap\&SS]
  {10.1007/s10509-007-9296-z}, \href
  {http://adsabs.harvard.edu/abs/2007Ap%26SS.308..625W} {308, 625}

\bibitem[\protect\citeauthoryear{{Yoshida} \& {Lee}}{{Yoshida} \&
  {Lee}}{2002}]{YoshidaLee2002}
{Yoshida} S.,  {Lee} U.,  2002, \mn@doi [Astron. Astrophys.]
  {10.1051/0004-6361:20021270}, \href
  {http://adsabs.harvard.edu/abs/2002A%26A...395..201Y} {395, 201}

\bibitem[\protect\citeauthoryear{{Zanotti}, {Morozova}  \& {Ahmedov}}{{Zanotti}
  et~al.}{2012}]{Zanotti2012}
{Zanotti} O.,  {Morozova} V.,   {Ahmedov} B.,  2012, \mn@doi [Astron.
  Astrophys.] {10.1051/0004-6361/201118380}, \href
  {http://adsabs.harvard.edu/abs/2012A%26A...540A.126Z} {540, A126}

\bibitem[\protect\citeauthoryear{{Zink}, {Lasky}  \& {Kokkotas}}{{Zink}
  et~al.}{2012}]{Zink2012}
{Zink} B.,  {Lasky} P.~D.,   {Kokkotas} K.~D.,  2012, \mn@doi [Phys. Rev. D]
  {10.1103/PhysRevD.85.024030}, \href
  {http://adsabs.harvard.edu/abs/2012PhRvD..85b4030Z} {85, 024030}

\makeatother
\end{thebibliography}
\input{manuscript.bbl}

\appendix

\section{Electromagnetic fields in the wave zone for higher order spheroidal
modes}
\label{em_wz_spheroidal}

When the oscillation mode is non-axisymmetric, the electromagnetic fields
produced by a spheroidal oscillation mode with $\ell'=2, m'=2$ have a
rather complicated form. To obtain such expressions, we start by noting
that for a radiating mode with $\ell=3$, the only nonzero coefficient is
given by $u_{32}$ and has explicit expression (\cf Section
\ref{spheroid_wz})
\begin{equation}
\label{u_32_sph_wz} u_{32}=-\frac{1}{15\sqrt{21}}
    \frac{\omega^3_{_\textrm{R}} R^4}{N_{_\textrm{R}}}B_{_0}
    \left(3\xi_{_\textrm{R}}f_{_\textrm{R}}-2\eta_{_\textrm{R}}h_{_\textrm{R}}\right) \cos\chi\,.
\end{equation}
Similarly, the quadrupolar outgoing radiation defined by $\ell=2$ has the
only nonzero coefficient given by $v_{22}=0$ and explicit form
\begin{equation}
\label{v_22_sph_wz} v_{22}=-\frac{1}{3}\sqrt{\frac{1}{6}}
    \frac{\omega^3_{_\textrm{R}} R^4}{N_{_\textrm{R}}}B_{_0}
    \eta_{_\textrm{R}}h_{_\textrm{R}} \cos\chi\,.
\end{equation}
Using these results, the multipolar electromagnetic fields
(\ref{sol_wz_mf1})--(\ref{sol_wz_ef3}) induced in the wave-zone by the
oscillation mode $\ell'=2, m'=2$ are then expressed as the real parts of
the following solutions for the magnetic field
\begin{eqnarray}
\label{wz_spheroid_2_1} && B^{\hat r} =
    \frac{1}{2\sqrt{30\rm{\pi}}}\frac{\omega^2_{_\textrm{R}}R^4}{N^2_{_\textrm{R}}r^2}
    B_{_0} \left(2\eta_{_\textrm{R}}h_{_\textrm{R}}-3\xi_{_\textrm{R}}f_{_\textrm{R}}\right)
    \sin^2\theta\cos\theta
        \cos\chi {\,{\rm e}^{{\rm i}[\omega (r-t)+2\phi]}} \,,
\\ \nonumber\\
\label{wz_spheroid_2_2} && B^{\hat\theta} =
    \frac{{\rm i}}{48}\frac{1}{\sqrt{30\rm{\pi}}}
    \frac{\omega^3_{_\textrm{R}}R^4}{N_{_\textrm{R}}r}
    B_{_0} \left[22\eta_{_\textrm{R}}h_{_\textrm{R}}-3\xi_{_\textrm{R}}f_{_\textrm{R}}+
    3\left(2\eta_{_\textrm{R}}h_{_\textrm{R}}-3\xi_{_\textrm{R}}f_{_\textrm{R}}\right)
    \cos 2\theta\right]
    \sin\theta\cos\chi {\,{\rm e}^{{\rm i}[\omega (r-t)+2\phi]}} \,,
\\ \nonumber\\
\label{wz_spheroid_2_3} && B^{\hat\phi} =
    -\frac{1}{24\sqrt{30\rm{\pi}}}
    \frac{\omega^3_{_\textrm{R}}R^4}{N_{_\textrm{R}}r}
    B_{_0} \left(7\eta_{_\textrm{R}}h_{_\textrm{R}}-3\xi_{_\textrm{R}}f_{_\textrm{R}}\right)
    \sin\theta\cos\theta
        \cos\chi {\,{\rm e}^{{\rm i}[\omega (r-t)+2\phi]}}  \,,
\end{eqnarray}
while for the electric field, they assume the form
\begin{eqnarray}
\label{wz_spheroid_2_4} && E^{\hat r} =  \frac{{\rm i}}{4}
    \sqrt{\frac{5}{6\rm{\pi}}}\frac{\omega^2_{_\textrm{R}}R^4}{N^2_{_\textrm{R}}r^2}
    B_{_0} \eta_{_\textrm{R}}h_{_\textrm{R}}
        \cos\chi {\,{\rm e}^{{\rm i}[\omega (r-t)+2\phi]}} \,,
\\ \nonumber\\
\label{wz_spheroid_2_5} && E^{\hat\theta} =
    -\frac{1}{24\sqrt{30\rm{\pi}}}
    \frac{\omega^3_{_\textrm{R}}R^4}{N_{_\textrm{R}}r}
    B_{_0} \left(7\eta_{_\textrm{R}}h_{_\textrm{R}}-3\xi_{_\textrm{R}}f_{_\textrm{R}}\right)
    \sin\theta\cos\theta
        \cos\chi {\,{\rm e}^{{\rm i}[\omega (r-t)+2\phi]}}  \,,
\\ \nonumber\\
\label{wz_spheroid_2_6} && E^{\hat\phi} =
    \frac{{\rm i}}{48}\frac{1}{\sqrt{30\rm{\pi}}}
    \frac{\omega^3_{_\textrm{R}}R^4}{N_{_\textrm{R}}r}
    B_{_0} \left[22\eta_{_\textrm{R}}h_{_\textrm{R}}-3\xi_{_\textrm{R}}f_{_\textrm{R}}+
    3\left(2\eta_{_\textrm{R}}h_{_\textrm{R}}-3\xi_{_\textrm{R}}f_{_\textrm{R}}\right)
    \cos 2\theta\right]
    \sin\theta\cos\chi {\,{\rm e}^{{\rm i}[\omega (r-t)+2\phi]}} \,.
\end{eqnarray}
%

\section{Electromagnetic fields in the wave zone for higher order toroidal
modes}
\label{em_wz_toroidal}

In similarity to what done in Appendix \ref{em_wz_spheroidal}, we next
list the explicit wave-zone expressions for the electromagnetic fields
radiated by a toroidal oscillations. We start with a mode with $\ell'=1,
m'=1$, in which case the components of the magnetic field are
\begin{eqnarray}
&& B^{\hat r} = \frac{3{\rm i}}{2\sqrt{6\rm{\pi}}}
    \frac{f_{_\textrm{R}}R^2}{N_{_\textrm{R}}^2r^2} B_{_0}
    \eta_{_\textrm{R}}\sin\theta\cos\chi  {\,{\rm e}^{{\rm i}[\omega (r-t) + \phi]}} ,
\\ \nonumber\\
&& B^{\hat\theta} = -\frac{3}{4\sqrt{6\rm{\pi}}}
    \frac{f_{_\textrm{R}}\omega_{_\textrm{R}}R^2}{N_{_\textrm{R}}r}
    \left(1+\frac{{\rm i}}{9}\omega^2_{_\textrm{R}}R^2\right) B_{_0}
        \eta_{_\textrm{R}}\cos\theta\cos\chi {\,{\rm e}^{{\rm i}[\omega (r-t) + \phi]}} \,,
\\ \nonumber\\
&& B^{\hat\phi} = -
    \frac{3{\rm i}}{4\sqrt{6\rm{\pi}}}
    \frac{f_{_\textrm{R}}\omega_{_\textrm{R}}R^2}{N_{_\textrm{R}}r}
    \left(1+\frac{{\rm i}}{9}\omega^2_{_\textrm{R}}R^2\cos 2\theta\right)
    B_{_0}\eta_{_\textrm{R}}\cos\chi {\,{\rm e}^{{\rm i}[\omega (r-t) + \phi]}} \,,
\end{eqnarray}
while the electric field is given by
\begin{eqnarray}
&& E^{\hat r} = -
    \frac{{\rm i}}{2\sqrt{6\rm{\pi}}}
    \frac{f_{_\textrm{R}}\omega^2_{_\textrm{R}}R^4}{N_{_\textrm{R}}^2r^2}
    B_{_0}
        \eta_{_\textrm{R}}\sin\theta\cos\theta\cos\chi
    {\,{\rm e}^{{\rm i}[\omega (r-t) + \phi]}} \,,
\\ \nonumber\\
&& E^{\hat\theta} = -
    \frac{3{\rm i}}{4\sqrt{6\rm{\pi}}}
    \frac{f_{_\textrm{R}}\omega_{_\textrm{R}}R^2}{N_{_\textrm{R}}r}
    \left(1+\frac{{\rm i}}{9}
    \omega^2_{_\textrm{R}}R^2\cos 2\theta\right) B_{_0}
        \eta_{_\textrm{R}}\cos\chi {\,{\rm e}^{{\rm i}[\omega (r-t) + \phi]}} \,,
\\ \nonumber\\
&& E^{\hat\phi} =  \frac{3}{4\sqrt{6\rm{\pi}}}
    \frac{f_{_\textrm{R}}\omega_{_\textrm{R}}R^2}{N_{_\textrm{R}}r}
    \left(1+\frac{{\rm i}}{9}\omega^2_{_\textrm{R}}R^2\right) B_{_0}
        \eta_{_\textrm{R}}\cos\theta\cos\chi {\,{\rm e}^{{\rm i}[\omega (r-t) + \phi]}} \,,
\end{eqnarray}
where the non-vanishing integration constants are given by
\begin{equation}
\label{v_u_tor_wz_11} v_{21}=\frac{{\rm i}}{3\sqrt{30}}
    \frac{f_{_\textrm{R}}\omega^3_{_\textrm{R}} R^4}{N_{_\textrm{R}}}B_{_0}
    \eta_{_\textrm{R}} \cos\chi\,, \qquad
u_{11}=\frac{{\rm i}}{\sqrt{2}}
    \frac{f_{_\textrm{R}}\omega_{_\textrm{R}} R^2}{N_{_\textrm{R}}}B_{_0}
    \eta_{_\textrm{R}} \cos\chi\,.
\end{equation}

When the mode has $\ell'=m'=2$ (\ie the prototypical $r$ mode discussed
when considering gravitational-wave losses) the non-vanishing integration
constants will take the form [see also \citet{Ho2000} for a Newtonian
  treatment of the electromagnetic emission from a star subject to
  $r$-mode oscillations]
\begin{equation}
\label{v_u_tor_wz_22} v_{32}=\frac{{\rm i}}{15\sqrt{21}}
    \frac{f_{_\textrm{R}}\omega^4_{_\textrm{R}} R^5}{N_{_\textrm{R}}}B_{_0}
    \eta_{_\textrm{R}} \cos\chi\,, \qquad
u_{22}=\frac{{\rm i}}{3}\sqrt{\frac{2}{3}}
    \frac{f_{_\textrm{R}}\omega^2_{_\textrm{R}} R^3}{N_{_\textrm{R}}}B_{_0}
    \eta_{_\textrm{R}} \cos\chi\,,
\end{equation}
and consequently generate the following magnetic fields in the wave zone
\begin{eqnarray}
&& B^{\hat r}= -\frac{1}{2}\sqrt{\frac{5}{6\rm{\pi}}}
    \frac{f_{_\textrm{R}}\omega_{_\textrm{R}}R^3}{N_{_\textrm{R}}^2r^2} B_{_0}
    \eta_{_\textrm{R}}\cos\chi\sin^2\theta
    {\,{\rm e}^{{\rm i}[\omega (r-t) + 2\phi]}} \,,
\\ \nonumber\\
&& B^{\hat\theta}= -\frac{{\rm i}}{12}\sqrt{\frac{5}{6\rm{\pi}}}
    \frac{f_{_\textrm{R}}\omega^2_{_\textrm{R}}R^3}{N_{_\textrm{R}}r} B_{_0}
     \eta_{_\textrm{R}}
    \left(1+\frac{{\rm i}}{10}\omega^2_{_\textrm{R}}R^2\right)
    \sin\theta\cos\theta\cos\chi
    {\,{\rm e}^{{\rm i}[\omega (r-t) + 2\phi]}} \,,
\\ \nonumber\\
&& B^{\hat\phi}= \frac{1}{12}\sqrt{\frac{5}{6\rm{\pi}}}
    \frac{f_{_\textrm{R}}\omega^2_{_\textrm{R}}R^3}{N_{_\textrm{R}}r} B_{_0}
     \eta_{_\textrm{R}}
    \left(1+\frac{{\rm i}}{40}\omega^2_{_\textrm{R}}R^2+
    \frac{3{\rm i}}{40}\omega^2_{_\textrm{R}}R^2\cos 2\theta\right)
    \sin\theta \cos\chi
    {\,{\rm e}^{{\rm i}[\omega (r-t) + 2\phi]}} \,,
\end{eqnarray}
while the electric fields are
\begin{eqnarray}
&& E^{\hat r} =
    \frac{1}{2\sqrt{30\rm{\pi}}}
    \frac{f_{_\textrm{R}}\omega^3_{_\textrm{R}}R^5}{N_{_\textrm{R}}^2r^2} B_{_0}
    \eta_{_\textrm{R}}\cos\chi\sin^2\theta\cos\theta
    {\,{\rm e}^{{\rm i}[\omega (r-t) + 2\phi]}} \,,
\\ \nonumber\\
&& E^{\hat\theta}=  \frac{1}{12}\sqrt{\frac{5}{6\rm{\pi}}}
    \frac{f_{_\textrm{R}}\omega^2_{_\textrm{R}}R^3}{N_{_\textrm{R}}r} B_{_0}
     \eta_{_\textrm{R}}
    \left(1+\frac{{\rm i}}{40}\omega^2_{_\textrm{R}}R^2+
    \frac{3{\rm i}}{40}\omega^2_{_\textrm{R}}R^2\cos 2\theta\right)
    \sin\theta \cos\chi
    {\,{\rm e}^{{\rm i}[\omega (r-t) + 2\phi]}} \,,
\\ \nonumber\\
&& E^{\hat\phi}=
    \frac{{\rm i}}{12}\sqrt{\frac{5}{6\rm{\pi}}}
    \frac{f_{_\textrm{R}}\omega^2_{_\textrm{R}}R^3}{N_{_\textrm{R}}r} B_{_0}
     \eta_{_\textrm{R}}
    \left(1+\frac{{\rm i}}{10}\omega^2_{_\textrm{R}}R^2\right)
    \sin\theta\cos\theta\cos\chi
    {\,{\rm e}^{{\rm i}[\omega (r-t) + 2\phi]}} \,.
\end{eqnarray}

\label{lastpage}

\end{document}